\documentclass[twocolumn]{aastex63}
\usepackage{graphicx}
\usepackage{amssymb}
\usepackage{amsmath}
\usepackage{epstopdf}
\usepackage{calc}
\newcommand{\bvec}[1]{{\ensuremath{\boldsymbol{#1}}}}
\newcommand{\T}{\ensuremath{\mathrm{T}}}

\shorttitle{Cosmic Star-Formation History}
\shortauthors{Matthews et al.}

\begin{document}

\pdfsuppresswarningpagegroup=1
\maxdeadcycles=1000

\title{Cosmic Star-Formation History Measured at 1.4\,GHz}

\correspondingauthor{Allison M. Matthews} \email{amm4ws@virginia.edu}

\author[0000-0002-6479-6242]{A.~M.~Matthews}
  \affiliation{Department of Astronomy,
  University of Virginia, Charlottesville, VA 22904, USA}
\affiliation{National Radio Astronomy Observatory, 520 Edgemont Road,
  Charlottesville, VA 22903, USA}

\author[0000-0003-4724-1939]{J.~J.~Condon}
\affiliation{National Radio Astronomy
  Observatory, 520 Edgemont Road, Charlottesville, VA 22903, USA}

\author[0000-0001-7363-6489]{W.~D.~Cotton}
  \affiliation{National Radio Astronomy
  Observatory, 520 Edgemont Road, Charlottesville, VA 22903, USA}

  \author[0000-0003-2716-9589]{T.~Mauch}
  \affiliation{South African Radio Astronomy
  Observatory (SARAO), 2 Fir Street, Black River Park, Observatory,
  7925, South Africa}

\begin{abstract}
We matched the 1.4\,GHz local luminosity functions of star-forming
galaxies (SFGs) and active galactic nuclei to the 1.4\,GHz
differential source counts from $0.25\,\mu\mathrm{Jy}$ to 25\,Jy using
combinations of luminosity and density evolution.  We present the
most robust and complete local far-infrared (FIR)/radio luminosity
correlation to date in a
volume-limited sample of $\approx 4.3 \times 10^3$ nearby
SFGs, finding that it is very tight but distinctly sub-linear:
$L_\mathrm{FIR} \propto L_\mathrm{1.4\,GHz}^{0.85}$.  If
the local FIR/radio correlation does not evolve,
the evolving 1.4\,GHz luminosity function of SFGs yields the
evolving star-formation rate density (SFRD) $\psi
(M_\odot\,\mathrm{year}^{-1}\,\mathrm{Mpc}^{-3}$) as a function of
time since the big bang. The SFRD measured at 1.4\,GHz grows rapidly
at early times, peaks at ``cosmic noon'' when $t \approx
3\,\mathrm{Gyr}$ and $z \approx 2$, and subsequently decays with an
$e$-folding time scale $\tau = 3.2\,\mathrm{Gyr}$.  This evolution is
similar to, but somewhat stronger than, SFRD evolution estimated from
UV and FIR data.

\end{abstract}


%%%%%%%%%%%%%%%%%%%%%%%%%%%%%%%%%%%%%%%%%%%%%%%%%%

%%%%%%%%%%%%%%%%% BODY OF PAPER %%%%%%%%%%%%%%%%%%

\section{Introduction}

Fundamental to our understanding of galaxy evolution, reionization of
the universe, and heavy element production is an evolutionary timeline
of the cosmic star formation rate density (SFRD).  In the 1990s, it
was first suggested that star-formation activity at redshift $z\sim1$
dwarfed that at $z\sim0$ \citep[e.g.][]{songaila94,ellis96,
lilly96}. In the decades since, star-forming galaxies have been
detected out to increasing redshifts, most recently $z\gtrsim10$
\citep[e.g.][] {coe13,oesch16}, well
within the reionization era.  Compilations of SFR measurements made
at various redshifts informs our understanding of SFRD evolution
\citep[see][for reviews of the topic]{hopkins06, madau14}.
Virtually all SFR diagnostics are
sensitive to massive stars only; an initial mass function (IMF) must
be assumed to tally the total stellar mass formed at a given time.
Possible variations in the IMF within and among galaxies and redshifts
remains a source of uncertainty.

Most extragalactic radio sources fainter than $S \approx
0.4\,\mathrm{mJy}$ at $1.4\,\mathrm{GHz}$ are distant star-forming
galaxies (SFGs), while stronger sources are primarily radio galaxies
or quasars powered by active galactic nuclei (AGNs)
\citep{prandoni01,smolcic08,condon12, vernstrom16}.
The 1.4\,GHz continuum emission from SFGs is a combination of
synchrotron radiation from electrons accelerated in the supernova
remnants of short-lived ($\tau \leq 3 \times 10^7 \,\mathrm{yr}$)
massive ($M > 8 M_\odot$) stars plus thermal bremsstrahlung from
H\textsc{ii} regions ionized and heated by even more massive stars
\citep{condon92}.  The cosmic-ray electrons responsible for the
synchrotron radiation dominating the 1.4\,GHz continuum emission
eventually diffuse throughout their host galaxy and cool on timescales
$\tau_{\rm cool} \sim 5$\,Myr \citep[for a spiral galaxy that stopped
forming stars after a single episode;][]{murphy08}. The combined
lifetimes of such massive stars with the cooling timescale of
cosmic-ray electrons are much less than the age of the universe, so
the radio continuum luminosities of SFGs depend only on their current
star-formation rates uncontaminated by older stellar populations.
Although the radio continuum luminosity is only a tiny fraction of the
total power emitted by massive stars, its tight correlation with the
energetically dominant far-infrared (FIR) emission from dust heated by
massive stars justifies the use of radio emission as a quantitative
tracer of star formation in galaxies \citep{condon92}.

Stars with masses $M\gtrsim 8 M_{\odot}$ emit primarily in the
ultraviolet (UV) continuum.  The rest frame wavelength range 1400\,\AA
\ to 1700\,\AA ~is accessible to ground-based telescopes for galaxies with
redshifts $z \gtrsim 1.4$, but the UV emission of nearby galaxies must
be measured either at longer UV wavelengths or from space.  The
contribution from longer-lived ($\tau\sim 250$\,Myr) radio-quiet stars
increases at longer UV wavelengths.  The biggest downside for UV
emission as a tracer of the SFR is dust obscuration.  At redshifts
$z\sim2$, dust attenuation measured via infrared/UV luminosity ratios
$L_{\rm IR}/L_{\rm UV}$ implies that $>$80\% of star formation is
obscured \citep{reddy12, howell10}, resulting in a small
UV contribution to the total SFRD.

The UV energy absorbed by dust grains is reemitted at mid-infrared
(MIR) and far-infrared (FIR) wavelengths, making MIR and FIR
luminosities practical SFR indicators (in the case of
  minimal contribution from diffuse dust).
Although FIR \citep[$42.5-122.5\,\mu\mathrm{m}$;][]{helou88} dust
extinction is low, the infrared spectrum (spanning $\lambda =
8-1000\,\mu$m) of a galaxy is complex. The fraction of UV luminosity
absorbed by dust depends on the metallicity and geometry of the dust
distribution, and the dust emission at wavelengths longer than
$\lambda \sim 100\,\mu$m in the source frame is powered primarily by
evolved stars \citep[e.g.][]{hirashita03,bendo10}.  At MIR
wavelengths, emission from warm dust is tightly correlated with star
formation, but polycyclic aromatic hydrocarbons (PAHs) complicate the
emission spectrum near $\lambda = 8\,\mu$m and active galactic nuclei
(AGNs) dilute these PAH features while also contributing
significantly to the 24$\,\mu$m continuum emission.
Luminous infrared galaxies (LIRGS) with $L_\mathrm{IR} > 10^{11}
L_\odot$ and ultra-luminous infrared galaxies (ULIRGS) with
$L_\mathrm{IR} > 10^{12} L_\odot$ are rare today but were responsible
for most of the luminosity density during the $z \sim 2$ ``cosmic
noon'' \citep{magnelli11} when most stars were formed.
The MIR emission due solely to star formation must be
  disentangled from the total MIR emission before converting to a
SFR to ensure a correct result.

The cosmic history of star formation can be constrained by a
combination of the 1.4\,GHz local luminosity function and the
differential numbers $n(S)dS$ of faint radio sources per steradian
with flux densities between $S$ and $S + dS$. A very low 1.4\,GHz
detection limit $S = 0.25\,\mu$Jy is needed to reach SFRs of
evolving ``normal'' galaxies like the Milky Way: $5\,
M_\odot\,\mathrm{yr}^{-1}$ at $z = 2$, $12\,M_\odot\,\mathrm{yr}^{-1}$
at $z = 3$, and $22\, M_\odot\,\mathrm{yr}^{-1}$ at $z = 4$ (assuming
a Salpeter IMF).  Thus the top continuum science goal of the proposed
Square Kilometre Array SKA1-MID is ``Measuring the Star-formation
History of the Universe'' using the proposed ``Ultra Deep Reference
Survey'' to count sources as faint as $S = 0.25\,\mu\mathrm{Jy}$ in a
solid angle $\Omega \approx 1\,\mathrm{deg}^2$ \citep{prandoni15}.
Recently \citet{condon19} measured the 1.4\,GHz local (${z < 0.1}$)
radio luminosity functions of SFGs and AGNs from sources
  in the 1.4\,GHz NRAO VLA Sky Survey \citep[NVSS]{condon98}
  cross-identified with 2MASX galaxies \citep{jarrett00}.
\citet{matthews21} determined accurate
1.4\,GHz brightness-weighted source counts $S^2 n(S)$ over the eight
decades of flux density between $S = 0.25 \,\mu\mathrm{Jy}$ and $S =
25\,\mathrm{Jy}$ using the very sensitive $\nu = 1.266\,\mathrm{GHz}$
MeerKAT DEEP2 sky image \citep{mauch20} for sources counts below
$S = 2.5\,\mathrm{mJy}$, and the 1.4~GHz NVSS catalog 
above $S = 2.5\,\mathrm{mJy}$ (see \cite{matthews21} for details).

In this paper we present the cosmic star-formation history derived
from only (1) the 1.4\,GHz local luminosity function,
(2) the local volume-limited FIR/radio correlation, and
(3) the 1.4\,GHz counts of
sources as faint as $S = 0.25\,\mu\mathrm{Jy}$.  We do not need to
``stack'' radio sources to achieve the required sensitivity, so we do
not depend on a complete sample of optically selected galaxies with
measured redshifts and do not discriminate against galaxies so
obscured by dust that they drop out of optical samples. The faintest
radio sources were detected statistically via their confusion $P(D)$
distribution, so we actually cannot optically identify them or measure
their redshifts.  Instead, their radio evolution is constrained
entirely by matching features in the local luminosity function to
features in the source counts.  This independent approach complements
the traditional methods reviewed by \citet{madau14}.

Section~\ref{sec:llfs} reviews and updates the 1.4\,GHz local
luminosity functions of SFGs and AGNs derived from a spectroscopically
complete sample of $\sim 10^4$ 2MASX \citep{jarrett00} galaxies
brighter than $k_{20fe} = +11.75$ at $\lambda = 2.2\,\mu$m and
stronger than $S = 2.5\,\mathrm{mJy}$ at $\nu = 1.4\,\mathrm{MHz}$.
Basic equations relating the evolving $1.4\,\mathrm{GHz}$ luminosity
functions and spectral-index distributions to the counts of distant
sources in the flat $\Lambda$CDM universe are introduced in
Section~\ref{sec:basics}. The non-evolving model source counts are
discussed in Section~\ref{sec:noevomod} to highlight the features that
evolutionary models must have to fit the 1.4\,GHz data.  Models for
the radio evolution of both AGNs and SFGs that successfully match
their evolving luminosity functions to the 1.4\,GHz source counts are
presented in Section~\ref{sec:evomods}.  We calculated an improved
local FIR/radio correlation using a large volume-limited sample of
SFGs in our 2MASX sample and found it to be a slightly nonlinear power
law: $L_\mathrm{FIR} \propto L_\mathrm{1.4\,GHz}^{0.85}$. We used this
local FIR/radio correlation to convert the evolving 1.4\,GHz SFG
luminosity functions into FIR star-formation rate densities (SFRDs)
and to make an independent estimate of the cosmic history of star
formation (Section~\ref{sec:sfhu}).  Section~\ref{sec:conclusions}
summarizes and evaluates these results.

Absolute quantities were calculated for the flat
$\Lambda$CDM universe with $H_0 = 70\,\mathrm{km\,s}^{-1}
\mathrm{\,Mpc}^{-1}$ and $\Omega_\mathrm{m} = 0.3$.  Our
spectral-index sign convention is $\alpha \equiv + d\, \ln S / d\,\ln
\nu$.  The \citet{salpeter55} IMF was used to calculate total
star-formation rates in terms of $M_\odot\,\mathrm{yr}^{-1}$.  These
rates should be multiplied by 0.61 for the \citet{chabrier03} IMF or
by 0.66 for the \citet{kroupa01} IMF.

\section{Local 1.4\,GHz Luminosity Functions}\label{sec:llfs}

The evolving spectral luminosity function $\rho(L_\nu \vert z) d
L_\nu$ specifies the comoving number density of sources at redshift
$z$ having absolute spectral luminosities $L_\nu$ to $L_\nu + d L_\nu$
at frequency $\nu$.  The corresponding density of sources per decade
of spectral luminosity is
\begin{equation}
  \rho_\mathrm{dex} (L_\nu \vert z) = \ln (10) L_\nu \rho(L_\nu \vert z)~.
\end{equation}
Sources with this luminosity function produce a comoving spectral
power density per decade of luminosity
\begin{equation}
  u_\mathrm{dex} (L_\nu \vert z) \equiv L_\nu \,
  \rho_\mathrm{dex} ( L_\nu \vert z)~.
\end{equation}
We call $u_\mathrm{dex}$ the energy-density function because spectral
power density has the same dimensions as energy density (SI units
$\mathrm{W\,Hz}^{-1} \mathrm{\,m}^{-3} = \mathrm{J\,m}^{-3}$).
Astronomically practical units for $u_\mathrm{dex}$ are
$\mathrm{W\,Hz}^{-1}\,\mathrm{dex}^{-1}\,\mathrm{Mpc}^{-3}$.

\begin{figure}[!htb]
  \includegraphics[width=0.5\textwidth,trim={2.2cm 9.5cm 5.5cm 9.5cm},clip]
  {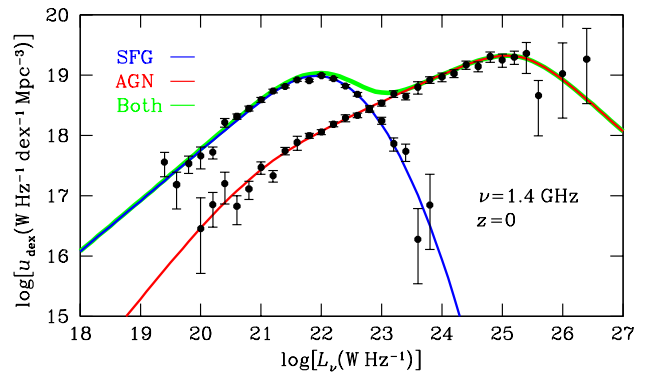}
  \caption{The measured local energy-density functions $u_\mathrm{dex}
    (L_\nu \vert 0)$ at $\nu = 1.4\,\mathrm{GHz}$ are shown by the
    black data points.  The red curve plots the
    Equation~\ref{eqn:udexagn} fit for AGNs, the blue curve plots
    Equation~\ref{eqn:udexsfg} for SFGs, and the wider green curve is the
    sum of both.
    \label{fig:udex0}
    }
\end{figure}

The local 1.4\,GHz energy-density functions $u_\mathrm{dex}(L_\nu
\vert 0)$ of radio sources powered primarily by active galactic nuclei
(AGNs) or by star-forming galaxies (SFGs) were determined separately
\citep{condon19} and are shown by the data points and error bars in
Figure~\ref{fig:udex0}.  These local 1.4\,GHz energy-density
  functions were determined from a large sample ($N \sim 1\times10^4$)
  of radio sources in the NVSS catalog covering $\Omega = 7.016$ sr of sky
  and cross-identified with $\lambda = 2.16\,\mu$m galaxies in the 2MASX.
  All 9517 sources have spectroscopic redshifts and radio sources powered
  primarily by AGNs were separated from those powered by SFGs using
  the following radio and infrared diagnostics: (1) an \textit{IRAS}
  FIR/NVSS 1.4\,GHz flux-density ratio $q < 1.8$, (2) a FIR spectral
  index $\alpha(25\,\mu\mathrm{m},\,60\,\mu\mathrm{m}) > -1.25$, (3)
  have \textit{WISE} colors $(W1-W2)>0.8$ for $(W2-W3) \geq 3.1$ and
  $(W1-W2)>(W2-W3-1.82)/1.6$ for $(W2-W3) < 3.1$, and (4) showed a radio
  morphology with multiple components (e.g. jets, lobes in the case of
  resolved NVSS sources).
  The median redshift (corrected for the local flow due to nearby
  galaxy clusters) is $\langle z\rangle\approx0.02$
  for the SFG sample ($N = 6699$) and $\langle z\rangle\approx0.04$ for
  the AGNs ($N = 2763$). For further details on the derivation of these
energy-density functions, we refer the reader to \cite{condon19}.
For computational convenience, we approximate
the AGN energy-density function by
\begin{equation}\label{eqn:udexagn}
  u_\mathrm{dex}(L_\nu \vert 0) = \frac {C_\mathrm{a} L_\nu}
  {(L_\nu/L^*_\mathrm{a})^\alpha + (L_\nu / 
    L^*_\mathrm{a})^\beta + (L_\nu /
    L_\mathrm{min})^\gamma}
\end{equation}
with comoving density factor $C_\mathrm{a} = 2.0 \times 10^{-6} \,
\mathrm{Mpc}^{-3}\allowbreak\,\mathrm{dex}^{-1}$, AGN high-luminosity
turnover spectral luminosity $L^*_\mathrm{a} = 2.0 \times 10^{25}
\,\mathrm{W\,Hz}^{-1}$, low-luminosity downturn luminosity
$L_\mathrm{min} = 1.0 \times 10^{11} \,\mathrm{W\,Hz}^{-1}$,
intermediate-luminosity power-law slope $\alpha = 0.55$,
high-luminosity power-law slope $\beta = 1.9$, and low-luminosity
downturn slope $\gamma = -0.25$.  This function is shown by the red
curve in Figure~\ref{fig:udex0}.  Its parameters are highly
correlated, so their values and their uncertainties have limited
physical significance.  However, the uncertainty of $\beta$ is
especially large because there are few AGNs with $L_\nu >
L^*_\mathrm{a}$ and redshifts $z < 0.1$.

% This figure was drawn by Count.dir/udex0.sm using data file udex0.OUT
% produced by udex0.f with agn log (xlstar) = 25.1 and betafit = +1.85

The tight FIR/radio correlation implies that the radio and FIR luminosity
functions of SFGs should have similar functional forms, so we followed the
standard form established by \citet{saunders90} for the $\lambda =
60\,\mu\mathrm{m}$ luminosity function to write
\begin{equation}\label{eqn:udexsfg}
  u_\mathrm{dex}(L_\nu \vert 0) = C_\mathrm{s} \biggl(
  \frac {L_\nu}{L^*_\mathrm{s}} \biggr)^{2-\alpha_\mathrm{s}}
  \hskip -.2cm \exp \biggl[ - \frac {1}{2\sigma^2} \log^2 \biggl( 1 +
    \frac {L_\nu} {L^*_\mathrm{s}} \biggr) \biggr]
\end{equation}
with comoving density factor $C_\mathrm{s} = 3.50 \times 10^{-3}\,
\mathrm{Mpc}^{-3}\allowbreak \,\mathrm{dex}^{-1}$, turnover spectral
luminosity $L^*_\mathrm{s} = 1.9 \times 10^{21} \,
\mathrm{W\,Hz}^{-1}$, $\alpha_\mathrm{s} = 1.162$ for low-luminosity
power-law slope $(2-\alpha_\mathrm{s}) = +0.838$, and high-luminosity
Gaussian taper with rms width $\sigma = 0.558$.  Our value of
$L^*_\mathrm{s}$ is close to the $L_\nu = 2.5 \times 10^{21}
\,\mathrm{W\,Hz}^{-1}$ 1.4\,GHz spectral luminosity of the Milky Way
\citep{berkhuijsen84}.  The local energy-density function of SFGs is
plotted as the blue curve in Figure~\ref{fig:udex0}, and the sum of
the AGN and SFG local energy-density functions is indicated by the
wider green curve.

The accessible volumes and hence numbers of galaxies with low 1.4\,GHz
luminosities used to calculate the local energy-density functions are
limited primarily by the $S \approx 2.5\,\mathrm{mJy}$ sensitivity
limit of the NVSS catalog, so the statistical uncertainties in these
energy-density functions increase for AGNs below $L_\nu \sim 10^{21}
\, \mathrm{W\,Hz}^{-1}$ and for SFGs below $L_\nu \sim 10^{20}
\,\mathrm{W\,Hz}^{-1}$.

\section{Basic Equations}\label{sec:basics}

The differential source count $n(S) dS$ is the
number of sources per steradian with flux densities between $S$ and $S
+ dS$. Defining $\eta(S) d\log(S)$ as the number of sources per
steradian per $\log(S)$ and substituting $dS = S d\ln(S) = \ln(10) S
d\log(S) $ shows that $\ln(10) S^2 n(S) = S \eta (S)$ is the flux
density per steradian (a spectral brightness) per decade of flux
density.  Thus the Rayleigh-Jeans sky brightness temperature $d
T_\mathrm{b}$ per decade of flux density contributed by sources is
\begin{equation}\label{eqn:tb}
\Biggl[ \frac {d \,T_\mathrm{b}} {d \log(S)}\Biggr] =
\Biggl[ \frac {\ln(10)\,c^2} { 2 k_\mathrm{B} \nu^2} \Biggr]  S^2 n(S)~,
\end{equation}
where $k_\mathrm{B} \approx 1.38 \times 10^{-23}\mathrm{\,J\,K}^{-1}$.
We call $S^2 n(S)$ the brightness-weighted differential source count
to distinguish it from the traditional static-Euclidean weighted
count $S^{5/2} n(S)$.

In a flat $\Lambda$CDM universe, the total brightness-weighted count
at frequency $\nu$ of sources with spectral index $\alpha$ can be
written as the integral of $u_\mathrm{dex}(L_\nu \vert z)$ over
redshift \citep{condon18}:
\begin{equation}\label{eqn:cms2n}
  S^2 n(S) = \frac {D_{H_0}} {4 \pi \ln(10)}
  \int_0^\infty u_\mathrm{dex} (L_\nu \vert z)
  \biggl[ \frac {(1+z)^{\alpha - 1}} {E(z)} \biggr] \,dz ~,
\end{equation}
where $D_{H_0} \equiv c / H_0$ is the Hubble distance, $L_\nu = 4 \pi
D_\mathrm{C}^2 (1+z)^{1-\alpha} S$, $D_\mathrm{C}$ is the comoving
distance to the source, and $E(z) = [\Omega_\mathrm{m}
  (1+z)^3 + \Omega_\Lambda + \Omega_\mathrm{r} (1+z)^4]^{1/2}$.

It is instructive to rewrite Equation~\ref{eqn:cms2n} in terms of
lookback time $t_\mathrm{L}(z)$ by substituting the relations
\citep{condon18}
\begin{equation}
  d D_\mathrm{C} = D_{H_0} \frac {dz}{E(z)} = (1 + z) c \,d t_\mathrm{L}
\end{equation}
to yield
\begin{equation}\label{eqn:tls2n}
  S^2 n(S) = \frac {c} {4 \pi \ln(10)}
  \int_0^{t_\mathrm{L}(\infty)} u_\mathrm{dex}
  (L_\nu \vert z) (1+z)^\alpha dt_\mathrm{L}\,,
\end{equation}
where $t_\mathrm{L}(z = \infty) \approx 0.964 H_0^{-1} \approx
13.47\,\mathrm{Gyr}$ is the current age of the universe.
Equation~\ref{eqn:tls2n} shows that the sources in any narrow range
$\Delta t_\mathrm{L}$ of lookback time near redshift $z(t_\mathrm{L})$
contribute
\begin{equation}\label{eqn:bincount}
  \Delta [S^2 n(S)] \propto u_\mathrm{dex}(L_\nu \vert z)
  (1+z)^\alpha \Delta [t_\mathrm{L}(z)]
\end{equation}
to the brightness-weighted source count.  Thus in a plot of $S^2 n(S)$
versus $\log(S)$, the contribution to $S^2 n(S)$ by sources in each
narrow time range $\Delta t_\mathrm{L}(z)$ mimics the evolving
energy-density function attenuated by the factor $(1+z)^\alpha$.

Both the AGN and SFG source populations span a range of spectral
indices $\alpha$ characterized by their redshift-dependent normalized
spectral-index distributions $p(\alpha \vert z)$, so a more accurate
version of Equation~\ref{eqn:cms2n} is
\begin{eqnarray}\label{eqn:s2n}
  S^2 n(S) = \frac {D_{H_0}} {4 \pi \ln(10)} \times \hskip 3cm \nonumber \\
  \int_{-\infty}^\infty \biggl\{
  \int_0^\infty u_\mathrm{dex} (L_\nu \vert z) p(\alpha \vert z)
  \biggl[\frac {(1+z)^{\alpha - 1}}{E(z)}\biggr] dz \biggr\} d\alpha\,.~~
\end{eqnarray}

% See Count.dir/Notes.pdf from 2020 Apr 23
% for notes on spectral-index distributions
% This figure was drawn by Count.dir/spindex.sm using data file spindex.OUT
% produced by spindex.f.
\begin{figure}[!ht]
  \includegraphics[width=0.5\textwidth,trim={2.5cm 12.9cm 7.5cm 8.6cm},clip]
    {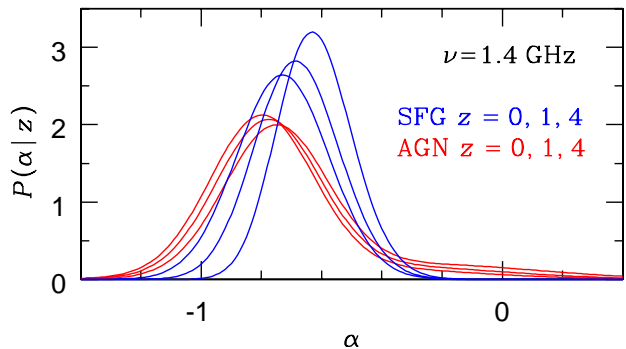}
  \caption{The 1.4~GHz normalized spectral-index distributions
    $P(\alpha \vert z)$ of SFGs (blue curves) and AGNs (red curves) 
    for sources at redshifts $z = 0,\,1$, and 4 (left to right).
    \label{fig:spindex}
    }
\end{figure}

The 1.4\,GHz spectral-index distribution of nearby AGNs can by
approximated by the the sum of two Gaussians representing the
steep-spectrum and flat-spectrum source populations \citep{condon84}:
\begin{eqnarray}
  p_\mathrm{a}(\alpha \vert 0) =
  \biggl( \frac{A_\mathrm{steep}}{\sqrt{2 \pi} \sigma_\mathrm{steep}} \biggr)
  \exp
  \biggl[ - \frac {(\alpha - \bar\alpha_\mathrm{steep})^2}
    {2 \sigma_\mathrm{steep}^2} \biggr] + \nonumber \\
  \biggl( \frac{A_\mathrm{flat}}{\sqrt{2 \pi} \sigma_\mathrm{flat}} \biggr)
  \exp
  \biggl[ - \frac {(\alpha - \bar\alpha_\mathrm{flat})^2}
    {2 \sigma_\mathrm{flat}^2} \biggr] \hskip 0.42cm 
\end{eqnarray}
with $A_\mathrm{steeo} = 0.86$, $\sigma_\mathrm{steep} = 0.17$,
$\bar\alpha_\mathrm{steep} = -0.8$ and $A_\mathrm{flat} = 1 - A_\mathrm{steep}
= 0.14$, $\sigma_\mathrm{flat} = 0.38$, $\bar\alpha_\mathrm{flat} = -0.5$.

The 1.4\,GHz spectral-index distribution of nearby SFGs can be
represented by a single population, but each SFG has two
spectral components---a nonthermal component with a Gaussian
spectral-index distribution characterized by mean spectral index
$\bar\alpha_\mathrm{n} \approx -0.8$ and rms width $\sigma_\mathrm{n}
\approx 0.17$ plus a thermal component with spectral index
$\alpha_\mathrm{t} \approx -0.1$.
At frequency $\nu$ in the source frame, the
nonthermal/thermal flux-density ratio $S_\mathrm{t}$ is
\citep{condon90}
\begin{equation}
  \frac {S_\mathrm{n}} {S_\mathrm{t}} \approx 10
  \biggl( \frac {\nu}{\mathrm{GHz}} \biggr)^{\alpha_\mathrm{n} + 0.1}\,.
\end{equation}
For $\nu_0= 1.4\,\mathrm{GHz}$ observations,
$S_\mathrm{n}/S_\mathrm{t}$ declines with redshift in the observed
  frame from 8 for galaxies
at $z = 0$ to 2.56 at $z = 4$.  Because nonthermal emission is always
dominant, the mean SFG spectral indices increase only slightly, from
$\langle\alpha\rangle \approx -0.72$ at $z = 0$ to
$\langle\alpha\rangle \approx -0.60$ at $z = 4$.

We have assumed that the locally measured spectral-index distributions
do not evolve in the source rest frame.  Even so, the observed $\nu_0 =
1.4\,\mathrm{GHz}$ spectral-index distributions of sources at redshift
$z$ are actually the spectral-index distributions of sources selected
at the higher frequency $\nu = (1+z) \nu_0$ in the source rest frame and
are biased toward ``flatter'' spectra with higher $\alpha$
\citep[see][appendix]{condon84}.  The expected 1.4\,GHz spectral-index
distributions of AGNs and SFGs at redshifts $z = 0, \,1\,$, and 4 are
compared in Figure~\ref{fig:spindex}.  Small changes in these
spectral-index distributions (e.g., varying $\bar\alpha_\mathrm{n}$ by
$\pm 0.1$) actually have very little effect on the predicted source
counts and redshift distributions.

\section{The Non-evolving Model}\label{sec:noevomod}

We integrated Equation~\ref{eqn:s2n} numerically to calculate $S^2
n(S)$ for the non-evolving model defined by $u_\mathrm{dex} (L_\nu
\vert z) = u_\mathrm{dex}(L_\nu \vert 0)$.  In order to show the
contributions to $S^2 n(S)$ from sources seen at different lookback
times $t_\mathrm{L}$, we broke the integration over $z$ into 13
redshift ranges corresponding to the 13 eons of lookback time $0 <
t_\mathrm{L} \mathrm{(Gyr)} < 1$, $1 < t_\mathrm{L} \mathrm{(Gyr)} <
2$, $2 < t_\mathrm{L} \mathrm{(Gyr)} < 3$, \dots, $12 < t_\mathrm{L}
\mathrm{(Gyr)} < 13$.  These lookback times and redshifts are listed
in Table~\ref{tab:ztl}.
 
% Data for this table are from Count.dir/countmod.f, declarations of
% zmin and zmax

\begin{deluxetable}{c c}[!ht]
  \tablecaption{Lookback Times and Redshifts \label{tab:ztl}}
  \tablehead{
    \colhead{$t_\mathrm{L}$\,(Gyr)} & Redshift $z$ 
  }
\startdata
  0 & 0.000 \\
  1 & 0.076 \\
  2 & 0.160 \\
  3 & 0.256 \\
  4 & 0.366 \\
  5 & 0.494 \\
  6 & 0.648 \\
  7 & 0.835 \\
  8 & 1.075 \\
  9 & 1.395 \\
  \llap{1}0 & 1.855 \\
  \llap{1}1 & 2.602 \\
  \llap{1}2 & 4.111 \\
  \llap{1}3 & 9.977 \\
\enddata
\end{deluxetable}  

% This figure showing no-evolution counts is the file
% Count.dir/countMOD0.pdf produced by Count.dir/countmod.sm with
% the 2020 Oct 30 updated fit to the AGN LLF

\begin{figure*}[!hbt]
  \includegraphics[width=\textwidth,trim={1.5cm 9.5cm 1.5cm 8cm},clip]
    {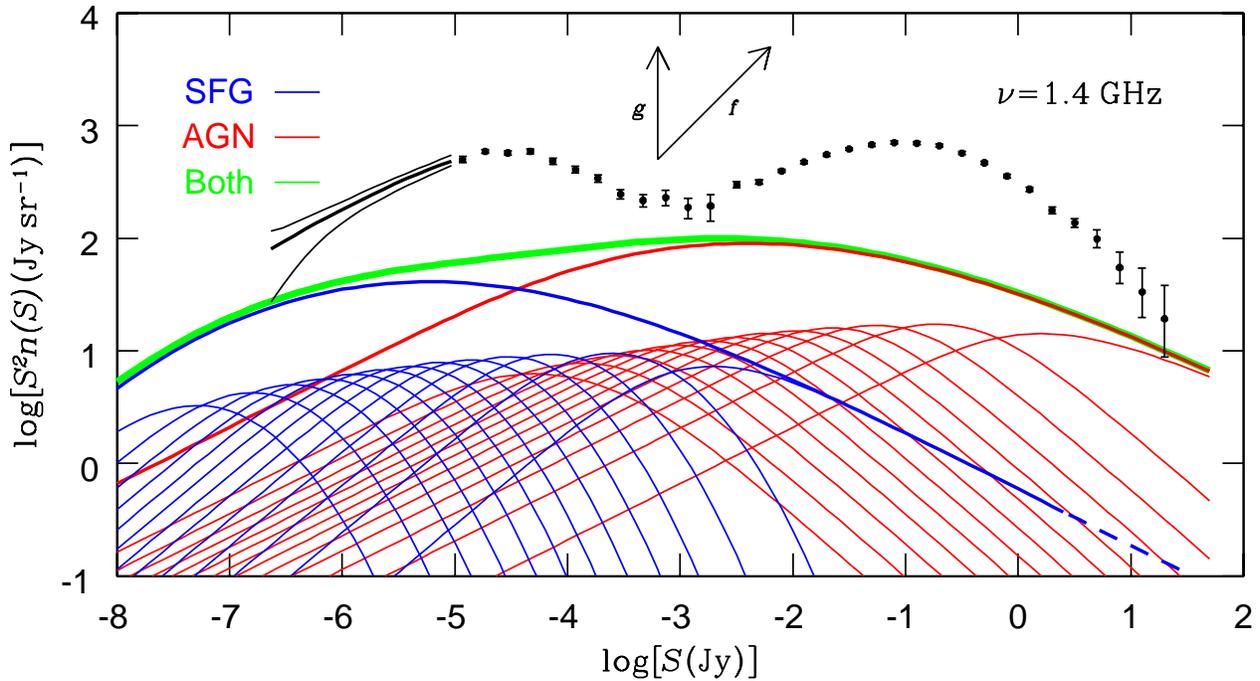}
  \caption{The heavy black curve spanning $-6.6 < \log[S\mathrm{(Jy)}]
    < -5.0$ marks the 1.4 GHz source count $S^2 n(S)$ determined from
    the DEEP2 confusion $P(D)$ distribution \citep{matthews21}, and
    the light black curves bound its rms uncertainties.  The data
    points and their rms error bars are the 1.4~GHz DEEP2 direct
    source counts in the range $-5 < \log[S\mathrm{(Jy)}] < -2.7$\,Jy
    and the NVSS counts for $\log[S\mathrm{(Jy)}] > -2.6$.  Below the
    data are curves showing the calculated source counts $S^2 n(S)$
    with no evolution.  The thick red curve is the total AGN count,
    the thick blue curve is the total SFG count, and the wider green curve
    is their sum, the total non-evolving model source count.  The dashed
    extrapolation of the thick blue curve shows the static Euclidean
    slope $d [\log[S^2 n(S)] / d [\log(S)] = -0.5$ expected in the limit of
    high flux densities where only low-redshift sources exist. The
    count contributions by sources in the 13 ranges of lookback time
    are shown by the lighter red and blue curves.  From right to left,
    the lookback time ranges are $t_\mathrm{L} =$ 0--1\,Gyr,
    1--2\,Gyr, \dots, 12--13\,Gyr.  The arrows labeled $f$ and $g$
    indicate how much a light curve covering a limited time range would be shifted by $f =
    10 \times$ luminosity evolution or by $g = 10 \times$ density
    evolution in that time range.
      \label{fig:noevo}
      }
\end{figure*}

With no evolution of the measured local energy-density functions,
Equation~\ref{eqn:s2n} gives the brightness-weighted source counts
$S^2 n(S)$ plotted in Figure~\ref{fig:noevo}.  The 13 thin red curves
from right to left are the AGN contributions from the 13 eons $0 <
t_\mathrm{L} \mathrm{(Gyr)} < 1$ through $12 < t_\mathrm{L}
\mathrm{(Gyr)} < 13$, and the thick red curve is the total
contribution from all AGNs with $t_\mathrm{L} < 13 \,\mathrm{Gyr}$ ($z
< 9.977$).  The blue curves show the analogous SFG contributions.  The
wider green curve is their sum, the total source count $S^2 n(S)$ for
the non-evolving model.

At the highest flux densities the model counts indicated by the heavy
red, blue, and green curves all must approach the static Euclidean
limit of nearby sources $n(S) = k S^{-5/2}$ whose plotted slope is $d
\log[S^2 n(S)] / d \log(S) = -1/2$.  The static Euclidean number of
sources per steradian stronger than $S$ is $N(>S) = (2 k /
3)S^{-3/2}$, and the thick curves have been truncated at the flux
densities above which they are statistically ill-defined because they
imply only one source in the entire sky: $N(>S) = (4\pi)^{-1}$.
Non-evolving sources in every $\Delta t_\mathrm{L} = 1\,\mathrm{Gyr}$
range of lookback time emitted the same total energy, so their
contributions to the sky brightness temperature $T_\mathrm{b}$ are
nearly equal, reduced moderately by the $(1+z)^\alpha$ attenuation
factor in Equation~\ref{eqn:bincount}.

Not only does the wide green curve lie well below the observed source
count, it is too smooth because the thick red and blue model curves
produced by summing over lookback times are much broader than the
peaks in the observed brightness-weighted source counts.  The arrows
labeled $f$ and $g$ in Figure~\ref{fig:noevo} indicate the effects of
$10 \times$ luminosity or density evolution, respectively, on counts
covering a limited time range.  Luminosity evolution moves the model
curves diagonally upward and to the right while density evolution
moves them straight up.  The peak in the thick blue curve lies
diagonally below and left of the SFG peak in the actual source counts
near $\log[S(\mathrm{Jy})] = -4.5$, so nearly pure luminosity
evolution should match the observed SFG counts.  The peak in the thick
red curve must move to the right more than it must move up to match
the AGN source-count peak near $\log[S\mathrm{(Jy)]} = -1$, suggesting
stronger luminosity evolution and negative density evolution.  The
strongest evolution should be confined to a narrow range of early
times in order to bunch up the light red and blue curves and narrow
the peaks of the heavy red and blue curves.  Just making the local
energy-density functions (Figure~\ref{fig:udex0}) match these features
of the brightness-weighted source counts (Figure~\ref{fig:noevo})
strongly constrains the redshift dependences of the luminosity
evolution $f(z)$ and density evolution $g(z)$, without depending on
measured redshifts for individual sources.

\section{Evolutionary Models}\label{sec:evomods}

We considered so-called backward evolutionary models (
  e.g. a local luminosity function is evolved backwards to match the observed
  source counts) in which the forms of the AGN and
SFG energy-density functions on a log-log plot
(Figure~\ref{fig:udex0}) do not change, but both populations may
evolve independently in both luminosity and density.  Pure luminosity
evolution $f(z)$ shifts the curves in Figure~\ref{fig:udex0}
diagonally upward and to the right, while pure density evolution
$g(z)$ shifts them vertically.  Then for each source population
\begin{equation}\label{eqn:udexevoz}
  u_\mathrm{dex} (L_\nu \vert z) =
  g(z)\, u_\mathrm{dex}\biggl[\frac{L_\nu} {f(z)} \vert 0\biggr]\,.
\end{equation}
%JC: explantion added per referee request
For any combination of luminosity evolution $f(z)$ and density evolution $g(z)$,
the total comoving spectral power density produced by
galaxies of all luminosities at redshift $z$ is proportional to 
the product $f(z)g(z)$.  Thus
\begin{eqnarray}
 U_\mathrm{SFG}(z) \equiv  \int_{-\infty} ^\infty u_\mathrm{dex}(L_\nu \vert z)\,
  d \log(L_\nu) = \nonumber \\ 
  f(z)g(z) \int_{-\infty}^\infty u_\mathrm{dex}(L_\nu \vert 0)\, d \log(L_\nu)~.
\end{eqnarray}
The total 1.4\,GHz spectral luminosity density produced by SFGs today is
\citep{condon19}
\begin{equation}\label{eqn:Usfg}
U_\mathrm{SFG}(0) =
(1.54 \pm 0.20) \times 10^{19} \,\mathrm{W\,Hz}^{-1}\,\mathrm{Mpc}^{-3}.
\end{equation}
  
Evolution is often described by functions of the observable source
redshift $z$, but for a specific cosmological model (e.g., our
$\Lambda$CDM model with $H_0 = 70\,\mathrm{km\,s}^{-1}
\mathrm{\,Mpc}^{-1}$ and $\Omega_\mathrm{m} = 0.3)$, $z$ can be used
to calculate the world time $t$ elapsed between the big bang and when the
source emitted the radiation we see today.  We prefer to describe
evolution in terms of $t$ because (1) the evolution experienced by a
source depends only on the time $t$ of emission, while $z$ also
depends on the time of the observation and (2) $z$ is a very nonlinear
measure of time (Table~\ref{tab:ztl}), so that equations expressing
evolution as a function of $z$ present a distorted picture of the
time scales involved.  Thus we chose to describe evolution as
\begin{equation}\label{eqn:udexevot}
  u_\mathrm{dex} (L_\nu \vert t) = g(t) \, u_\mathrm{dex}
  \biggl[\frac{L_\nu} {f(t)} \vert 0\biggl]~,
\end{equation}
subject to the boundary condition ${f(0) \cdot g(0) = 0}$ at the big
bang and $f(t_0) = g(t_0) = 1$ at the present time $t_0 \approx 13.47
\mathrm{\,Gyr}$.  

Models with strong luminosity evolution predict the existence at high
redshifts of extremely luminous AGNs that should have been observed,
but were not. \citet{peacock85} suggested cutting off the high end of
the luminosity function at all redshifts: $\rho(L_\nu) \propto
\exp(-L_\nu / L_\mathrm{c})$.  We applied this exponential cutoff with
$L_\mathrm{c} = 10^{29} \,\mathrm{W\,Hz}^{-1}$.

We model the luminosity and density evolution of both SFGs and AGNs as
the product of factors representing their rise at early times and
later exponential decay.  The rise is modeled in terms of
\begin{equation}
  \mathrm{erf}(t) \equiv  \frac{2}{\pi^{1/2}}
  \int_0^t e^{-x^2} dx,
\end{equation}
the $S$-shaped error function that increases from
$\mathrm{erf}(-\infty) = -1$ through $\mathrm{erf}(0) = 0$ to
$\mathrm{erf}(+\infty) = +1$, where $t$ is the age of the galaxy
in Gyr.  An exponential decay at larger $t$ is justified empirically by
the UV and IR data points from \cite{madau14} that fall on a nearly
straight line for $t>4$\,Gyr.

We represent luminosity evolution $f(t)$ and density evolution
$g(t)$ by the forms:
\begin{eqnarray}\label{eq:evoformf}
  f(t) &=& \left\{0.5\left[\mathrm{erf}\left(\frac{t-t_f}{\tau_f}\right) + 1\right]\right\}
  \,\left[\exp\left(\frac{t_0-t}{\tau_1}\right)\right]\\\label{eq:evoformg}
  g(t) &=& \left\{0.5\left[\mathrm{erf}\left(\frac{t-t_g}{\tau_g}\right) + 1\right]\right\}
  \,\left[\exp\left(\frac{t_0-t}{\tau_2}\right)\right],\qquad
\end{eqnarray}
where $t_f$ and $t_g$ are the midpoint times in Gyr of the turn-on phase
of luminosity and density evolution, $\tau_f$ and $\tau_g$ are the time 
scales of the turn-on, and $\tau_1$ and $\tau_2$
are the time scales in Gyr of luminosity and density decay.  In
Equations~\ref{eq:evoformf} and \ref{eq:evoformg}
$f$ and $g$ are the products of the turn-on
function in curly braces and the decay function in square brackets,
and both factors approach unity at $t = t_0 \approx
13.47\,\mathrm{Gyr}$.  The six free parameters are the time scales
$\tau_f$, $\tau_g$, $\tau_1$, and $\tau_2$ and the midpoint times
$t_f$ and $t_g$. Our choice of functional form has
the following desirable features: (1) it is continuous and smoothly
varying, (2) the asymptotic rise of the error function to $+1$ at
large $t$ makes the rise and decay factors cleanly separable, and (3)
the parameters have real physical meanings that can be compared with
independent measurements or theories (e.g. the rise time scale
$\tau_{\mathrm{r}}$ must agree with theoretical predictions for the
minimum time needed for the first galaxies to assemble).

AGNs dominate the 1.4\,GHz differential source counts (black points in
Figure \ref{fig:counts}) for all $\log[S\mathrm{(Jy)}] > -3.4$ ($S >
0.4 \,\mathrm{mJy}$) and SFGs outnumber AGNs at lower flux densities.
Nearly all of the AGNs contributing to $S^2 n(S)$ below
$\log[S\mathrm{(Jy)}] \approx -2$ come from the low-luminosity ($L_\nu
< L^*$) end of the AGN energy-density function
(Equation~\ref{eqn:udexagn}), which is nearly a power law.  Thus
\emph{all} AGN evolutionary models consistent with
Equations~\ref{eqn:udexevoz} or \ref{eqn:udexevot} and that match $S^2
n(S)$ for $\log[S\mathrm{(Jy)}] > -2$ must yield similar power-law
count contributions throughout the flux-density range dominated by
SFGs, as shown by the heavy red line in Figure~\ref{fig:counts}.
Consequently, uncertainties in the counts attributed to AGNs
  have little effect on the modeled SFG counts for $\log[S(\mathrm{Jy})]
    < -3.4$.

%JC: I slightly rephrased parts of the following paragraph:
It is mathematically inappropriate to judge the goodness-of-fit of our
evolutionary functions through a traditional non-linear least-squares
fit (or similar) of the predicted to the observed source counts
because the source-counts in adjacent flux-density bins are strongly
correlated and thus violate the independence assumption behind these
fitting methods. We used Gaussian processes
  \citep{rasmussen06} to accomodate these correlations and derive
  evolutionary models with appropriately conservative uncertainties in
  the fitted parameters. There are 6 free parameters in Equations
  \ref{eq:evoformf} and \ref{eq:evoformg} for both SFGs and AGNs (a
  total of 12). Because we assumed no late-time density evolution of
  SFGs, $\tau_\mathrm{2,SFG}$ is infinite, so we simultaneously fit only
  11 free parements in Equations \ref{eq:evoformf} and
  \ref{eq:evoformg}, plus two more parameters that describe the
  covariance between data points, using the affine-invariant Monte
  Carlo Markov Chain (MCMC) code \textit{emcee} \citep{emcee}. We
  assumed uniform priors for all parameters and enforced a slightly
  relaxed boundary condition $f(0) \cdot g(0) < 0.25 \approx 0$.  More
  details of our incorporation of Gaussian processes, the parameter
  contours resulting from the MCMC fitting, and marginalized posterior
  distributions can be found in Appendix \ref{sec:appendixa}.  

\subsection{AGN radio evolution}

%% This figure is the file Count.dir/MODA9S12count.pdf produced by
% Count.dir/countmod.sm for madelagn = 9, modelsfg = 12 on Nov. 1

\begin{figure*}[!ht]
  \includegraphics[width=\textwidth,trim={1.5cm 9.5cm 1.5cm
      8cm},clip]{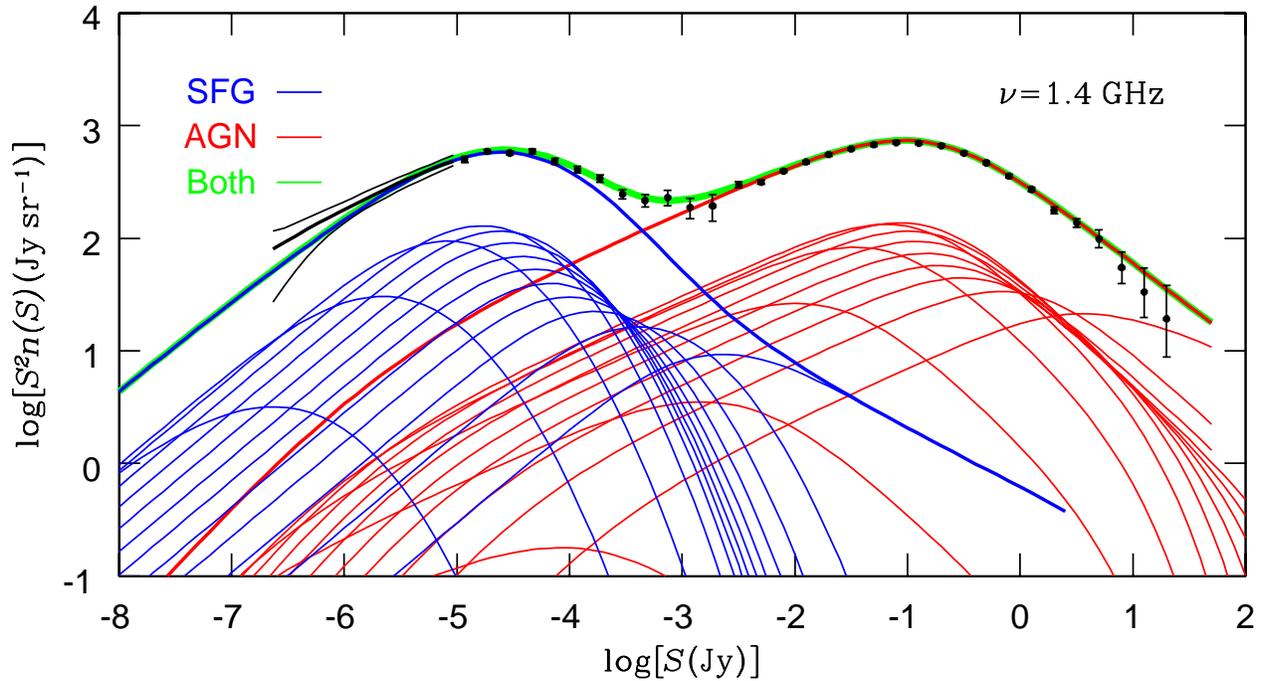}
  \caption{The 1.4 GHz differential source counts between $0.25
    \,\mu\mathrm{Jy}$ and $25\,\mathrm{Jy}$ are shown with the
    brightness-weighted normalization $S^2 n(S)$.  The thick black
    curve spanning $-6.6 < \log[S\mathrm{(Jy)} < -5$ is based on the
    DEEP2 confusion $P(D)$ distribution.  The data points with error
    bars show the 1.4~GHz DEEP2 source counts between in the range
    $-5 < \log[S\mathrm{(Jy)} < -2.6$\,Jy and the NVSS counts for
    $\log[S\mathrm{(Jy)} > -2.6$.  The thick curves show the total
    model counts for AGNs (red), SFGs (blue), and their sum
    (green).  The counts contributed by sources in the 13 ranges
    of lookback time are shown by the lighter red and blue
    curves.  From right to left, the time ranges are 0--1\,Gyr,
    1--2\,Gyr, \dots 12--13\,Gyr.  
      \label{fig:counts}
      }
\end{figure*}

\begin{figure*}[!ht]
  \includegraphics[width=0.45\textwidth,trim={4.cm 9.5cm 6.5cm
      9cm},clip]{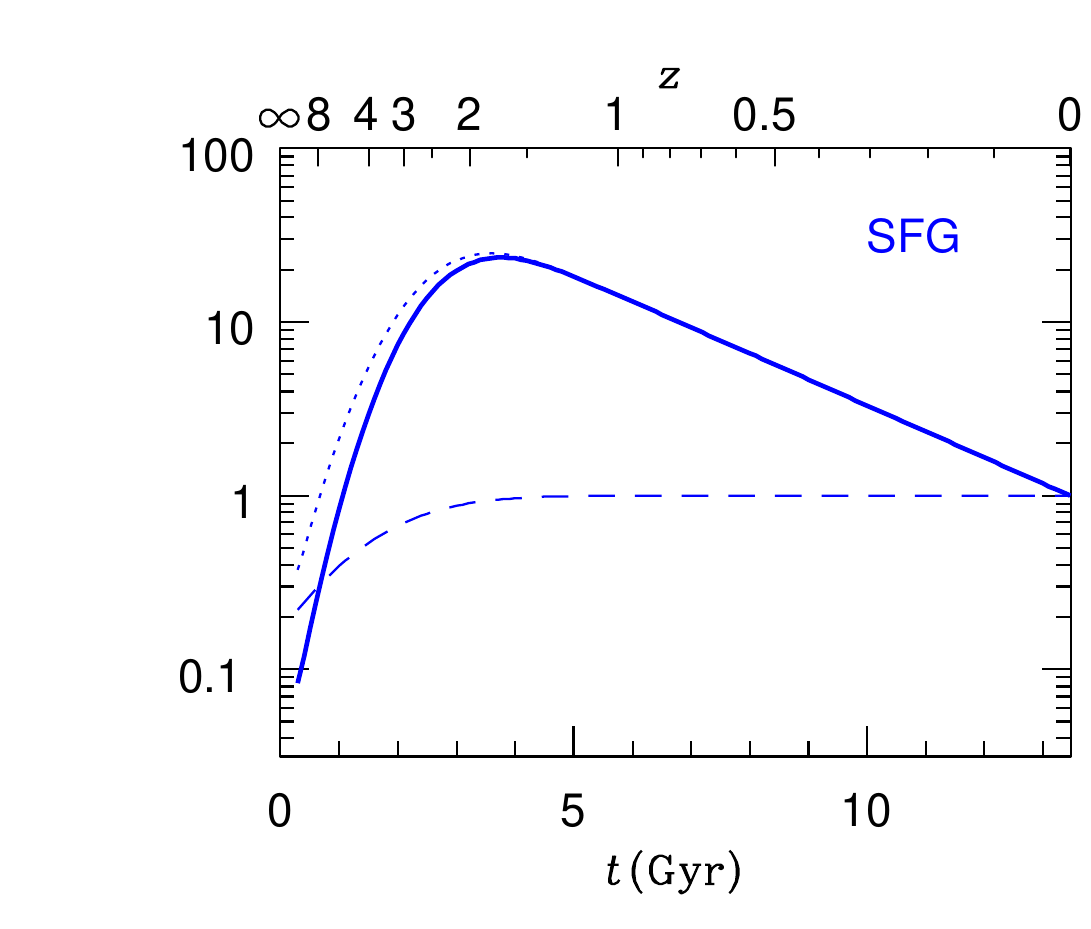}
  \includegraphics[width=0.45\textwidth,trim={4.cm 9.5cm 6.5cm
      9cm},clip]{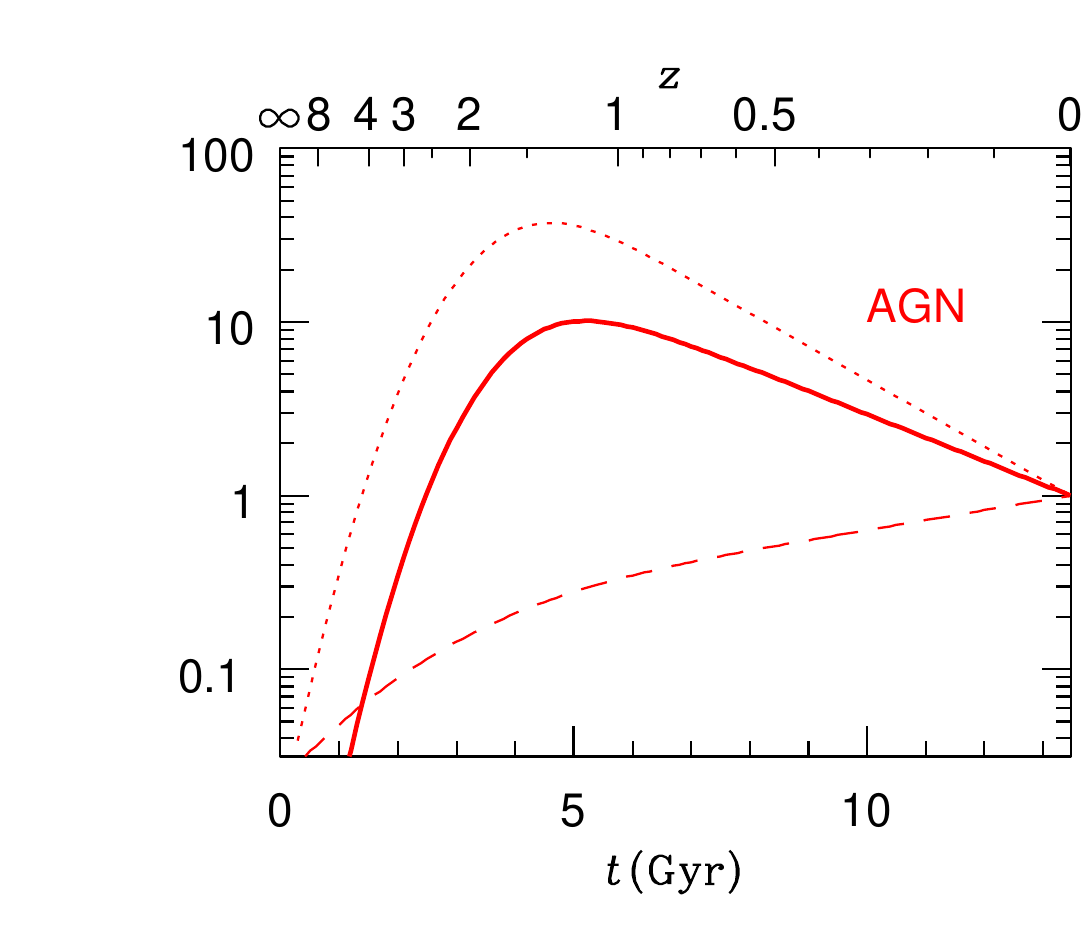}  
  \caption{The amounts of radio luminosity evolution $f$ (dotted
    curves), density evolution $g$ (dashed curves), and their products
    $fg$ (solid curves) best fitting the observed source counts are
    shown separately for SFGs (blue) and AGNs (red).
      \label{fig:evomod}
      }
\end{figure*}

As expected, pure luminosity evolution
($g = 1$) cannot match the observed sharp peak in $S^2 n(S)$ near
$\log[S\mathrm{(Jy)}] = -1$, so we had to supplement luminosity
evolution with negative density evolution ($g < 1$). Our best model
for AGN evolution has:
\begin{equation}\label{eqn:fa}
  f_\mathrm{a} = \left\{0.5\left[\mathrm{erf} \left(\frac{t - 3.97}{1.41}\right) + 1\right]\right\}\,
  \left[\exp\left(\frac{t_0-t}{2.26}\right)\right]\qquad
\end{equation}
and
%JC: I put the minus sign in the denominator to make tau2 negative, as listed in Table 2.
\begin{eqnarray}\label{eqn:ga}
  g_\mathrm{a} =  \left\{0.5\left[\mathrm{erf} \left(\frac{t - 2.59}{3.31}\right) + 1\right]\right\}\,
  \left[\exp\left(\frac{t_0-t}{-7.62}\right)\right]\qquad
\end{eqnarray}
where $t$ is the time in Gyr since the big bang. The negative decay time
scale $\tau_2 = -7.62$\,Gyr indicates a slow exponential \emph{growth}
in AGN density at late times. Uncertainties of
  the derived parameters and their correlations are shown in Appendix
  \ref{sec:appendixa}.
% and the quantities in
% braces represent the initial growth at small times $t$ with a rise time
% scale of 2.1\,Gyr and 1.5\,Gyr in luminosity and density, respectively.
% In Equation~\ref{eqn:fa} the
% quantity in square brackets represents exponential luminosity decay
% with time scale $2.25\,\mathrm{Gyr}$.  The square brackets in
% Equation~\ref{eqn:ga} give a late-time exponential growth in density
% with time scale $8\,\mathrm{Gyr}$ and ensure $g_\mathrm{a} = 0$ at $t
% = 0$. At $t = t_0 = 13.47\,\mathrm{Gyr}$, $f_\mathrm{a} = g_\mathrm{a}
% = 1$ by definition.
Figure~\ref{fig:evomod} plots $f_\mathrm{a}(t)$
and $g_\mathrm{a}(t)$ separately as dotted and dashed red curves,
respectively.  The total AGN spectral luminosity density
\begin{equation}
  U_\mathrm{AGN}(t) \equiv
  \int_{-\infty}^\infty u_\mathrm{dex}(L_\nu \vert t)
  \,d \log(L_\nu)
\end{equation}
is proportional to the product $f_\mathrm{a}(t)g_\mathrm{a}(t)$ shown
by the continuous red curve in Figure~\ref{fig:evomod} and
$U_\mathrm{AGN}(t_0) = (4.23 \pm 0.78) \times 10^{19}
\,\mathrm{W\,Hz}^{-1}\,\mathrm{Mpc}^{-3}$ at $\nu = 1.4\,\mathrm{GHz}$
\citep{condon19}.

Recall from Section~\ref{sec:llfs} and Figure~\ref{fig:udex0} that the
local energy-density function of AGNs is well determined down to
$\log[L_\nu\mathrm{(W\,Hz}^{-1})] \sim 21$, which is four decades
below the peak spectral luminosity $\log[L_\nu\mathrm{(W\,Hz}^{-1})]
\approx 25$.  Thus  the AGN contribution to the brightness-weighted
counts (Figure~\ref{fig:counts}) peaking at $\log[S\mathrm{(Jy)}]
\approx -1$ is well determined down to $\log[S\mathrm{(Jy)}] \sim
-5$, where the AGN contribution is only $\sim 3$\% of the
SFG contribution.  Any uncertainty in the numbers of fainter AGNs
is too small to affect either the total source counts or the
counts of SFGs.

\subsection{SFG radio evolution} \label{sec:SFGevo}

The radio evolution of SFGs at 1.4\,GHz is best fit by
\begin{equation}\label{eqn:fs}
  f_\mathrm{s}  =  \left\{0.5\left[\mathrm{erf}\left(\frac{t - 2.74}{1.30}\right) + 1\right]\right\}
  \left[\exp\left(\frac{t_0-t}{2.90}\right) \right]
\end{equation}
and
\begin{equation}\label{eqn:gs}
  g_\mathrm{s}  =  \left\{0.5\left[\mathrm{erf}\left(\frac{t - 1.38}{1.99}\right) + 1\right]\right\}\,,
  \quad\quad\qquad~
\end{equation}
where $t$ is the time in Gyr since the big bang and $\mathrm{erf}(t)$
is the error function.  For both luminosity evolution $f_\mathrm{s}$ and
density evolution $g_\mathrm{s}$, the quantities in braces specify the
$S$-shaped growth at early times.  At later times, the luminosity
%JC: I updated the SFG exponential time scale from 3.0 to 2.9 Gyr to match your MCMC result
evolution decays exponentially on a 2.9\,Gyr $e$-folding time scale,
and there is no density evolution.  These evolution functions
$f_\mathrm{s}$, $g_\mathrm{s}$, and their product
$f_\mathrm{s}g_\mathrm{s}$ are shown by the blue curves in
Figure~\ref{fig:evomod}.  Today $U_\mathrm{SFG}(t_0) = (1.54 \pm 0.20)
\times 10^{19} \,\mathrm{W\,Hz}^{-1}\,\mathrm{Mpc}^{-3}$
(Equation~\ref{eqn:Usfg}), and the resulting fits to the observed
faint-source counts are shown by the heavy blue (SFGs only) and
green (all sources) curves in Figure~\ref{fig:counts}.

\begin{figure}[!htb]
  \includegraphics[width=0.5\textwidth,trim={5cm 9.5cm 4.5cm
      9.7cm},clip]{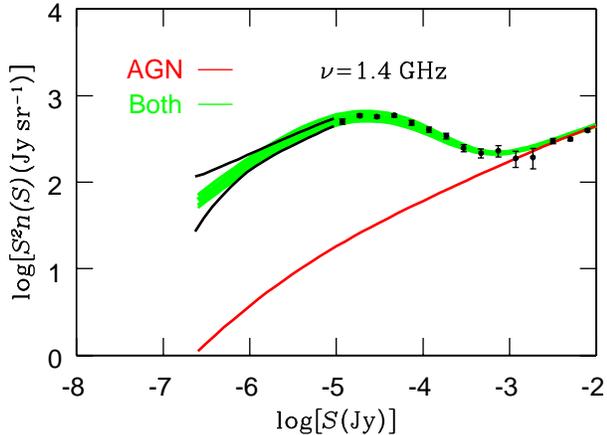}
  \caption{The broad green curve spans the range of total source
    counts bounded by the minimum and maximum SFG evolution models
    (Equations~\ref{eqn:fsmin} through \ref{eqn:gsmax}). The best-fit
    AGN counts are shown by the red curve.  The black data points with
    error bars are the DEEP2 and NVSS discrete source counts, and the
    black curves are the upper and lower limits of the $P(D)$ counts.
    \label{fig:minmax}
    }
\end{figure}

%JC: I slightly reworded the next paragraph.
To estimate the overall uncertainty in SFG evolution, we selected
  those MCMC parameter vectors yielding log-likelihood values
  in the highest 68\% of all samples. In the selected subsample, the minimum amount
of SFG evolution consistent with the 1.4\,GHz source counts is
\begin{equation}\label{eqn:fsmin}
  f_\mathrm{s}  =  \left\{0.5\left[\mathrm{erf}\left(\frac{t - 3.10}{1.12}\right) + 1\right]\right\}  \,
  \left[\exp\left(\frac{t_0-t}{3.04}\right) \right]
\end{equation}
\begin{equation}\label{eqn:gsmin}
  g_\mathrm{s}  =  \left\{0.5\left[\mathrm{erf}\left(\frac{t - 1.79}{0.42}\right) + 1\right]\right\}\,
  \qquad\qquad\quad\quad\quad
\end{equation}
and the maximum is
\begin{equation}\label{eqn:fsmax}
  f_\mathrm{s}  =  \left\{0.5\left[\mathrm{erf}\left(\frac{t - 2.51}{2.50}\right) + 1\right]\right\}  \,
  \left[\exp\left(\frac{t_0-t}{2.76}\right) \right]
\end{equation}
\begin{equation}\label{eqn:gsmax}
  g_\mathrm{s}  = \left\{0.5\left[\mathrm{erf}\left(\frac{t - 1.79}{0.97}\right) + 1\right]\right\}\,.
  \qquad\qquad\quad~~~ 
\end{equation}
The broad green curve in Figure~\ref{fig:minmax} shows the range of
counts bounded by these minimum and maximum evolution equations.
We stress that although the individual
  parameters describing the luminosity and density evolution have
  larger uncertainties (Appendix \ref{sec:appendixa}), the resulting
  total evolutionary curves remain consistent because the parameters
  are correlated. This ensures that the
  resulting implications for the star-formation history of the
  universe are stable.

When calculating far-ultraviolet (FUV) and FIR luminosity densities of
SFGs, \citet{madau14} truncated their luminosity functions below $0.03
L^*_\mathrm{s}$ (their equation 14).  As a test, we tried truncating
our 1.4\,GHz SFG luminosity function below $0.03 L^*_\mathrm{s}
\approx 6 \times 10^{19} \,\mathrm{W\,Hz}^{-1}$. The predicted counts
above $S \approx 0.25 \,\mu\mathrm{Jy}$ remained well within the green
curve in Figure~\ref{fig:minmax} and $\log[S^2 n(S)]$ fell by only
$0.08$ at $S\log[S(\mathrm{Jy})] = -8$.

\subsection{Sky brightness contributed by extragalactic sources at 1.4\,GHz}
\label{sec:tb}

\begin{figure}[!htb]
  \includegraphics[width=0.55\textwidth,trim={1.7cm 9.5cm 6.3cm
      10.8cm},clip]{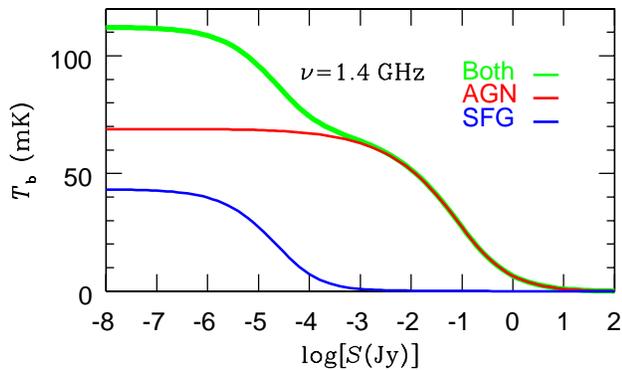}
  \caption{Model contributions to the 1.4\,GHz sky brightness
    temperature $T_\mathrm{b}$ from AGNs (red), SFGs (blue), and the
    sum of both (green) by sources with flux densities $>S$.
    \label{fig:tbcum}
    }
\end{figure}

Integrating Equation~\ref{eqn:tb} yields the
Rayleigh-Jeans sky brightness temperature
contributed by all sources stronger than $S_0$:
\begin{equation}\label{eq:deltatb}
  T_{\rm b}(>S_0) = \left[\frac{\ln(10)c^2}{2k_{\rm B}\nu^2}\right]
  \int_{\log S_0}^\infty S^2n(S) d(\log S).
\end{equation}
As shown in Figure~\ref{fig:tbcum}, the model that best fits the
brightness-weighted source counts $S^2 n(S)$ of AGNs adds
%JC: I ran your best new MCMC model through my source-count program
% including the full spectral-index distributions, and it increased
% the AGN contribution to the sky brightness from 68 to 69 mK.
% The SFG contribution is still 43 mK.
$T_\mathrm{b} \approx 69\,\mathrm{mK}$ to the Rayleigh-Jeans sky
brightness temperature at 1.4\,GHz, half of which comes from sources
stronger than $\log[S\mathrm{(Jy)}] = -1.2$ and 99\% from sources with
$\log[S\mathrm{(Jy)}] > -4.8$.  The acceptable range of SFG model
counts adds $T_\mathrm{b} = 43 \pm 6 \,\mathrm{mK}$ to the background,
of which $\approx 96$\% is resolved into sources stronger than $S =
0.25\,\mu\mathrm{Jy}$.  By integrating the backward evolutionary
  model for SFG out to increasing redshifts, we determine that
half of the total SFG background is produced
by sources having redshifts $z < 0.93 \pm 0.10$.

\begin{figure}[!htb]
  \includegraphics[width=0.48\textwidth,trim={1.3cm 9.5cm 6.5cm
      9.5cm},clip]{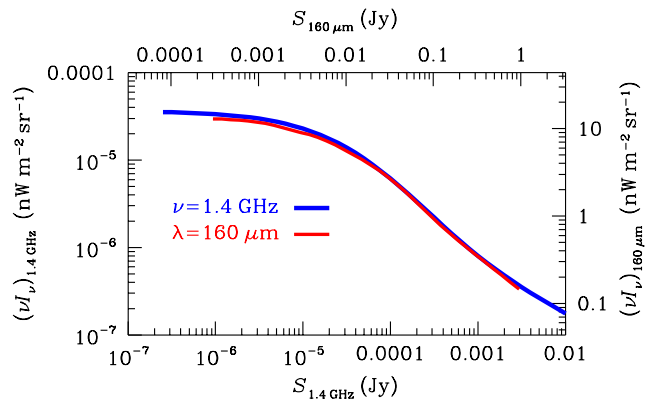}
  \caption{The $\nu = 1.4$\,GHz sky brightness temperature contributed
    by SFGs (blue curve in Figure~\ref{fig:tbcum}) was converted to
    the cumulative sky brightness $\nu I_\nu$ contributed by sources
    with flux densities $ > S_\mathrm{1.4\,GHz}$ and is shown by the
    blue curve (against the lower abscissa and left ordinate).  The
    red curve (against the upper abscissa and right ordinate) shows
    the $\lambda = 160\,\mu$m brightness contributed by sources
    stronger than $S_{160\,\mu\mathrm{m}}$ \citep{berta11}.  The
    curves overlap as shown when $S_{160\,\mu\mathrm{m}} /
    S_\mathrm{1.4\,GHz} = 310$ (Appendix~\ref{sec:appendixb}).
    \label{fig:tbcum2}
    }
\end{figure}

The $\nu = 1.4$\,GHz sky brightness temperature $T_\mathrm{b}$
produced by SFGs (blue curve in Figure~\ref{fig:tbcum}) was converted
to the 1.4\,GHz sky brightness $\nu I_\nu = 2 k_\mathrm{B} T_
\mathrm{b} \nu^3 c^{-2}$ in units of $\mathrm{nW \,
  m}^{-2}\,\mathrm{sr}^{-1}$ and is shown by the blue curve plotted
against the lower abscissa and left ordinate of
Figure~\ref{fig:tbcum2}. The $\lambda = 160\,\mu\mathrm{m}$ sky
brightness of faint FIR sources was measured by \citet{berta11} and is
shown by the red curve plotted against the upper abscissa and right
ordinate of Figure~\ref{fig:tbcum2}.  The left end of the red curve at
$S_{160\,\mu\mathrm{m}} = 0.3\,\mathrm{mJy}$ marks the sensitivity
limit of the \emph{Herschel} PACS $P(D)$ counts, and the right end
between $S_{160\,\mu\mathrm{m}} = 0.2 \,\mathrm{Jy}$ and 1\,Jy is the
static Euclidean extrapolation \citep{berta11}.  The upper abscissa
was shifted left by the expected mean flux-density ratio $\langle
S_{160\,\mu\mathrm{m}} / S_\mathrm{1.4\,GHz} \rangle \approx 310$ of
faint SFGs at median redshift $\langle z \rangle \approx 1$
\citep{condon19,berta11}, and the right ordinate for $\nu I_\nu$ was
shifted down by $4.16 \times 10^5$, the flux-density ratio multiplied
by the frequency ratio.  See Appendix~\ref{sec:appendixb} for the
derivation of these numbers.  The surprisingly good agreement of the
$\lambda = 160\,\mu\mathrm{m}$ and $\nu = 1.4\,\mathrm{GHz}$ SFG
backgrounds is reassuring evidence that (1) contamination of the SFG
population by radio-loud AGNs is small and (2) the local FIR/radio
correlation does not break down at redshifts $z \sim 1$.

The \emph{COBE} Far Infrared Absolute Spectrophotometer (FIRAS)
measured the total cosmic infrared background contributed by
all extragalactic sources to be $\nu I_\nu = 12.8 \pm 6.4
\,\mathrm{nW\,m}^{-2}\,\mathrm{sr}^{-1}$ at $\lambda =
160\,\mu\mathrm{m}$ \citep{fixsen98}, with zodiacal dust emission
causing most of the uncertainty.  If
$\langle S_{160\,\mu\mathrm{m}} / S_\mathrm{1.4\,GHz} \rangle \approx
310$, the corresponding 1.4\,GHz SFG background $\nu I_\nu = 3.1 \pm
1.5 \times 10^{-5} \,\mathrm{nW\,m}^{-2}\,\mathrm{sr}^{-1}$ is
consistent with the $\nu I_\nu \approx 3.5 \pm 0.5 \times 10^{-5}
\,\mathrm{nW\,m}^{-2}\,\mathrm{sr}^{-1}$ we obtained for SFGs stronger
than $S_\mathrm{1.4\,GHz} = 0.25\,\mu\mathrm{Jy}$.  Thus any
hypothetical ``new population'' of fainter radio sources bright enough
to produce the large extragalactic brightness at $\nu =
3.02\,\mathrm{GHz}$ reported by \citet{fixsen11} cannot obey the
FIR/radio correlation.

%% This figure is the file Count.dir/tbcum2.pdf produced by
% tbcum2.sm for model Count.dir/MODA9S12

\section{The Cosmic History of Star Formation}\label{sec:sfhu}

Section~\ref{sec:evomods} describes the \emph{radio} evolution needed
to match the local \emph{radio} energy-density function to the counts
of \emph{radio} sources associated with SFGs.  By themselves, these
quantities do not directly constrain the comoving SFRD $\psi(t)$
($M_\odot\,\mathrm{yr}^{-1}\,\mathrm{Mpc}^{-3}$).  To calculate the
evolving SFRD, we need a prescription relating the radio luminosities
of SFGs to their star-formation rates.  The radio continuum is an
energetically negligible tracer of star formation: the FIR/radio
luminosity ratio $\sim 4 \times 10^5$ of SFGs is comparable with the
elephant/mouse mass ratio.  Furthermore, most of the 1.4\,GHz emission
is synchrotron radiation whose luminosity depends on poorly known
quantities such as the interstellar magnetic field strength and
ambient radiation energy density.  It took the discovery of the
surprisingly strong empirical FIR/radio correlation in nearby galaxies
\citep{helou85} to convert radio continuum photometry of SFGs from a
hobby into a quantitative science.

\subsection{The linear FIR/radio correlation}

If the FIR/radio correlation is linear ($L_\mathrm{FIR} \propto
L_\mathrm{1.4\,GHz})$ and does not evolve, then only the local
FIR/radio flux-density ratio is needed to convert from radio
luminosity to star-formation rate.  That ratio is usually expressed in
terms of the dimensionless constant $q$ \citep{helou85}:
\begin{equation}\label{eqn:qdef}
  q \equiv \log \biggl[ \frac {\mathrm{FIR} /
      (3.75 \times 10^{12}\,\mathrm{Hz})}
    {S\mathrm{(1.4\,GHz)}}\biggr]~,
\end{equation}
where FIR is the flux between 42.5 and $122.5\,\mu$m in units
of W\,m$^{-2}$  estimated
from the \emph{IRAS} 60 and $100\,\mu$m flux densities in Jy
\begin{equation}\label{eqn:FIRdef}
  \mathrm{FIR} = 1.26 \times 10^{-14} [2.58 S(60\,\mu\mathrm{m}) +
    S(100\,\mu\mathrm{m})]\, 
\end{equation}
and $3.75 \times 10^{12}\,\mathrm{Hz}$ is the frequency corresponding
to the midpoint wavelength $\lambda = 80\,\mu\mathrm{m}$.  (Beware
that the FIR in Equation~\ref{eqn:FIRdef} is a flux with units of
W\,m$^{-2}$ so the numerator in Equation~\ref{eqn:qdef} is a flux
density with units of $\mathrm{W\,m}^{-2}\, \mathrm{Hz}^{-1}$.  Thus
either the denominator $S\mathrm(1.4\,\mathrm{GHz})$ should be
specified in units of $\mathrm{W\,m}^{-2}\, \mathrm{Hz}^{-1} =
10^{26}\,\mathrm{Jy}$ or, if $S\mathrm(1.4\,\mathrm{GHz})$ is
specified in Jy, the numerator should be multiplied by $10^{26}$.) For
the flux-limited \emph{IRAS} sample of galaxies with
$S(60\,\mu\mathrm{m}) > 2\, \mathrm{Jy}$, \citet{yun01} reported a
nearly linear FIR/radio correlation with scatter $\sigma = 0.26$ in
the $q$ values of individual galaxies and sample mean $\langle q
\rangle = 2.34 \pm 0.01$.

\begin{figure}[!htb]
  \includegraphics[width=0.46\textwidth,trim={5cm 9.5cm 6.5cm
      10.8cm},clip]{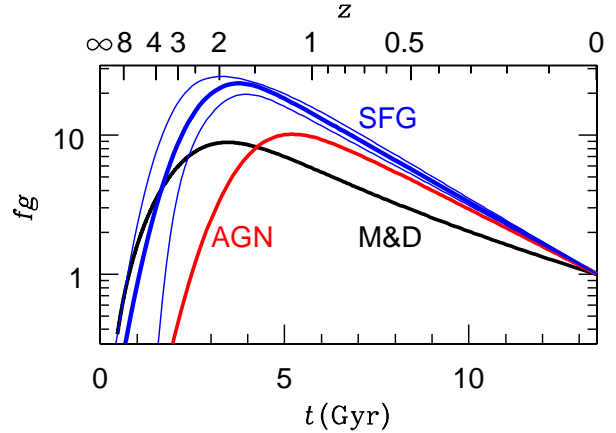}
  \caption{The thick blue curve, which is the same as the thick blue
    curve in Figure~\ref{fig:evomod}, shows the best-fit evolution of
    the radio SFRD if $\psi/\psi_0 = f_\mathrm{s} g_\mathrm{s}$ for
    SFGs, and the thin blue curves indicate the minimum and maximum
    amounts of evolution specified by Equations~\ref{eqn:fsmin}
    through \ref{eqn:gsmax}.  All are significantly higher than the
    black curve showing the evolution $\psi/\psi_0 = fg$ of the SFRD
    based on FUV and FIR data \citep[equation
      15]{madau14}.  The red curve is our best fit to the product
    $f_\mathrm{a}g_\mathrm{a}$ for AGN.  It is closer to the
    \citet{madau14} curve for stars, indicating comparable amounts of
    SFG and AGN evolution, but the AGN peak lags by $\gtrsim
    1\,\mathrm{Gyr}$. Abscissa: time $t$ in Gyr since the big
    bang. Ordinate: Normalized evolution $fg$.
    \label{fig:MODS9}
    }
\end{figure}

If the 1.4\,GHz spectral luminosities of SFGs are indeed
proportional to their star formation rates and the constant of
proportionality does not evolve, then the radio evolution of SFGs
implies SFRD evolution
\begin{equation}\label{eqn:psi}
  \frac {\psi(t)}{\psi_0} = f_\mathrm{s}(t) g_\mathrm{s}(t)\,,
\end{equation}
where $\psi_0 \equiv \psi(t_0)$ is the SFRD now.  The thick blue curve
in Figure~\ref{fig:MODS9} indicates the radio SFRD evolution based on
Equations~\ref{eqn:fs} and \ref{eqn:psi}.

% plot produced by Count.dir/evomodt.sm from data produced by
% Count.dir/evoplot.f

Using FIR and FUV data, \citet{madau14} estimated the
evolving SFRD and approximated it by the function
\begin{equation}\label{eqn:madau}
  \frac {\psi(z)}{\psi_0} \approx
  \frac { (1 + z)^{2.7} } {1 + [(1 + z) / 2.9]^{5.6}}
 \end{equation}
shown by the black curve in Figure~\ref{fig:MODS9}.  Both the blue and
black curves peak around the same ``cosmic noon'' near $t =
3\,\mathrm{Gyr}$, $z = 2$ and decline exponentially at later times,
but the radio estimate implies a significantly stronger overall
evolution of the SFRD.  If the FIR/radio correlation is linear and the
1.4\,GHz energy density function underwent luminosity evolution
specified by Equation~\ref{eqn:madau}, the predicted radio source
counts of SFGs would fall well below the observed counts. 
  Integrating the predicted radio source counts (see Equation
  \ref{eq:deltatb}) determines that
SFGs would contribute only $T_\mathrm{b} = 21 \,\mathrm{mK}$ to the sky
brightness temperature.

About half of the observed $43\,\mathrm{mK}$ SFG background is
produced by sources with $\log[S\mathrm{(Jy)}] > -4.8$ ($S >
16\,\mu\mathrm{Jy}$), nearly half by sources with $-6.6 <
\log[S\mathrm{(mJy)}] < -4.8$, and only $\sim 4$\% of the model SFG
background is produced by sources below our $P(D)$ count limit
$\log[S\mathrm{(Jy)]} = -6.6$ ($S = 0.25\,\mu\mathrm{Jy}$). Stronger
luminosity evolution and negative density evolution with fixed
$f_\mathrm{s}(z)g_\mathrm{s}(z)$ could fit the 1.4\,GHz source counts
above $\log[S\mathrm{(Jy)}] \sim -4.8$, but no separate adjustments of
luminosity evolution $f$ or density evolution $g$ consistent with a
given $\psi/\psi_0$ or product $fg$ can significantly change these SFG
contributions to $T_\mathrm{b}$ and match the counts between
$\log[S\mathrm{(Jy)]} = -6.6$ and $\log[S\mathrm{(Jy)]} = -4.8$.  We
conclude that the large difference between the radio and FUV/FIR SFRDs
cannot be avoided if the FIR/radio correlation is linear and
  does not evolve.  Thus authors who assume a linear FIR/radio
  correlation to model deep FIR and radio counts necessarily find that
  $q$ decreases with redshift; e.g., \citet{del17} used sensitive
  Jansky Very Large Array (VLA) and {\it Herschel} images to find $q
  \propto (1+z)^{-0.19 \pm 0.01}$ in the redshift range $0 < z < 6$.

\subsection{The nonlinear FIR/radio correlation}\label{sec:nonlinear}

The difference between SFRD evolution estimates based on our 1.4\,GHz
data and on the FUV/FIR data in \citet{madau14} can be reduced if the
FIR/radio correlation is sub-linear; that is, $x < 1$ in
$L_\mathrm{FIR} \propto L_\mathrm{1.4\,GHz}^x$.  To determine the
degree of nonlinearity, we measured the local $q$ (Equation~\ref{eqn:qdef}) as a
function of $\log[L\mathrm{(1.4\,GHz)}]$.  The $q$ distribution of
sources in a flux-limited sample is biased by the selection frequency;
thus the mean $\langle q \rangle$ in a FIR-selected sample is higher
than $\langle q \rangle$ in a radio-selected sample
\citep[see][appendix]{condon84}.  Such biases can be removed by
assigning to each source a weight inversely proportional to the
maximum volume $V_\mathrm{max}$ in which it could remain in the
sample, yielding the unbiased volume-limited distribution of $q$.

To measure the unbiased local distribution of $q$, we started with
  the large sample of NVSS sources stronger than $S = 2.5\,
  \mathrm{mJy}$ used in Section~\ref{sec:llfs} to determine the local
  radio luminosity function, but kept only the sources with $S \geq
  5\,\mathrm{mJy}$ to ensure that nearly all (98\%) of the sample SFGs
  were detected by \emph{IRAS} and have accurately measured values of
  $q$.  Although purely flux-limited samples of all radio sources with $S
  \geq 5\,\mathrm{mJy}$ are dominated by faint and distant ($\langle z
  \rangle \sim 1$) AGNs, our bright ($k_{20fe} < +11.75$) and thus
  local ($\langle z \rangle \sim 0.02$) sample is not, so the SFGs can
  be separated from the AGNs, as shown in Figure~\ref{fig:udex0}.

\begin{figure}[!ht]
% qofl5 includes only the S>5 mJy NVSS galaxies.
%See Count.dir/Notes.pdf 2020 Oct 16
  \includegraphics[width=0.49\textwidth,trim={1.9cm 12cm 7.5cm
      8cm},clip]{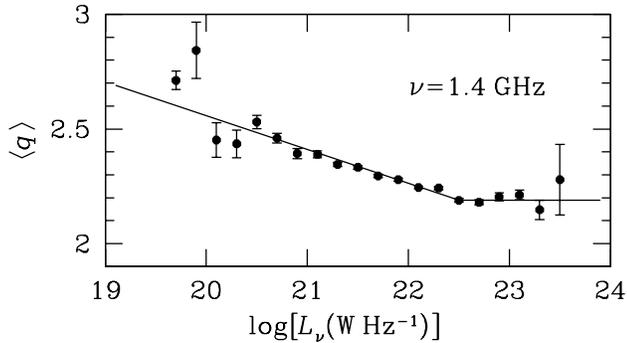}
  \caption{The logarithmic FIR/radio ratio parameter decreases as
    $d\langle q \rangle / d \log(L_\nu) = -0.147$ below $\log[L_\nu
      (\mathrm{W\,Hz}^{-1})] = 22.5$ and is a constant $\langle q
    \rangle = 2.19$ at higher luminosities.
      \label{fig:qofl}
      }
\end{figure}

This sample was divided into 1.4\,GHz luminosity bins of width $\Delta
\log(L_\nu) = 0.2$ centered on $\log(L_\nu) = 19.7$ through $23.5$. We
weighted the value of $q$ for each source by its $L_\nu /
V_\mathrm{max}$ ratio, where $V_\mathrm{max}$ is the smaller of its
$\lambda = 2.16\,\mu \mathrm{m}$ or $\nu = 1.4\,\mathrm{GHz}$ maximum
volumes, so that the overall weighted mean of the entire sample is an
unbiased measure of the volume-limited FIR/radio luminosity density
ratio.  Within each narrow radio luminosity bin, the rms scatter of
individual $q$ values is only $\sigma_q \approx 0.16$.  The weighted
means $\langle q \rangle$ and their rms uncertainties $\sigma_{\langle
  q \rangle}$ are plotted for all populated luminosity bins in
Figure~\ref{fig:qofl}.  The bent line in Figure~\ref{fig:qofl} shows
the fit
%from Count.dir/qofl5.sm
\begin{eqnarray}\label{eqn:qofl}
  \langle q \rangle  = 
  & 2.69 - 0.147[ \log(L_\nu) - 19.1] \nonumber 
  & ~~\mathrm{if} ~ \log(L_\nu) < 22.5 \quad \nonumber \\
  \langle q \rangle =
  & 2.19 ~~\mathrm{if} ~\log(L_\nu) \geq 22.5 \qquad
\end{eqnarray}
indicating a clearly sub-linear FIR/radio relation $L_\mathrm{FIR}
\propto L_\nu^{0.85}$ in the 1.4\,GHz luminosity range $\log(L_\nu) <
22.5$ that includes $>90$\% of nearby SFGs.  Sub-linearity implies
that FIR luminosity evolution, and hence SFRD evolution, is not as
strong as 1.4\,GHz evolution. The volume-limited average for nearby
SFGs of all luminosities is $\bar q = 2.30 \pm 0.01$.  These results
are quite stable, varying by $\sim 0.1$\% when the 2\% of galaxies
with only \emph{IRAS} upper limits are included or excluded.  To the
extent that the star-formation rates of galaxies are proportional to
their stellar masses $M_\star$ \citep{brinchmann04} (that is, there is
a ``main sequence'' of star-forming galaxies), the recent finding that
$dq / d \log(M_\star) = -0.148 \pm 0.013$ nearly independent of
redshift \citep{delvecchio20} is consistent with our sublinear local
FIR/radio correlation $dq / d \log(L_\nu) = -0.147$ and our assumption
that the FIR/radio correlation itself does not evolve with redshift.
While we know that the local FIR/radio correlation is sublinear
  and can fit our data with a non-evolving FIR/radio correlation, we
  cannot demonstrate that no such evolution exists.

For our evolutionary models, the FIR luminosities $L_\mathrm{FIR}$ of
individual SFGs at any redshift were estimated by inserting $\langle q
\rangle$ values from Equation~\ref{eqn:qofl} into
\begin{equation}
  L_\mathrm{FIR} =
  3.75 \times 10^{12}\,\mathrm{Hz}\cdot L_\mathrm{1.4\,GHz} \cdot
  10^{\langle q \rangle}~.
\end{equation}
The matching energy-density equation is
\begin{equation}
  \biggl[\frac {u_\mathrm{dex}(\mathrm{FIR})}{\mathrm{W\,Mpc}^{-3}}\biggr]
  = 3.75 \times 10^{12}\,\mathrm{Hz}\cdot
  \biggl[ \frac {u_\mathrm{dex}(\mathrm{1.4\,GHz})} {\mathrm{W\,Hz}^{-1}\,
  \mathrm{Mpc}^{-3}}\biggr] \cdot  10^q~.
\end{equation}

\subsection{Converting $L_\mathrm{FIR}$ to star-formation rates}

The total SFR associated with a given infrared
luminosity depends on the assumed initial mass function (IMF) and
stellar model spectra.  \citet{murphy11} assumed a \citet{kroupa01}
IMF and used the Starburst99 spectrum integrated over the infrared
(IR) band covering $8 < \lambda(\mu\mathrm{m}) < 1000$ to obtain
\begin{equation}\label{eqn:mureq4}
  \biggl( \frac {\mathrm{SFR}} {M_\odot \,\mathrm{yr}^{-1}} \biggr) =
    3.88 \times 10^{-37}
    \biggl( \frac {L_\mathrm{IR}} {\mathrm{W}} \biggr)~.
\end{equation}
The widely referenced conversion factor in table~1 of
  \citet{ken12} is based on this \citet{murphy11} value.
A Salpeter IMF \citet{salpeter55} has a larger fraction of low-mass
stars and implies that total SFRs including all stars in the mass range
$0.1 < M_\odot < 100$ are factor of $1 / 0.66 = 1.52$ higher for a
given $L_\mathrm{IR}$.  Most nearby SFGs have only measured FIR
luminosities, not IR luminosities.  \citet{bell03} compared the $q$
values for IR and FIR luminosities and found $\langle L_\mathrm{IR}
/ L_\mathrm{FIR} \rangle \approx \mathrm{dex}(2.64 - 2.36) \approx
1.91$, so for a \citet{kroupa01} IMF
\begin{equation}\label{eqn:sfrfir}
  \biggl( \frac {\mathrm{SFR}} {M_\odot \,\mathrm{yr}^{-1}} \biggr) =
    7.39 \times 10^{-37}
    \biggl( \frac {L_\mathrm{FIR}} {\mathrm{W}} \biggr)~.
\end{equation}

Combining these results and integrating over $\log L_\nu$ yields our
radio estimate of the evolving SFRD $\psi$ for a \citet{kroupa01} IMF
at any time $t$ :
\begin{eqnarray}\label{eqn:radiosfrd}
  \biggl[ \frac {\psi (t)}
    {M_\odot \,\mathrm{yr}^{-1} \, \mathrm{Mpc}^{-3}} \biggr] =
     7.39 \times 10^{-37} \,\cdot \,
    3.75 \times 10^{12}\mathrm{\,Hz}\, \cdot \nonumber \\
    \int \biggl[ \frac{u_\mathrm{dex}(L_\nu \vert t)
      \cdot 10^{\langle q(L_\nu) \rangle}} {\mathrm{W\,Hz}^{-1}\,
    \mathrm{Mpc}^{-3}}\biggr] d \log(L_\nu)\,,\qquad
\end{eqnarray}
where $L_\nu$ is the 1.4\,GHz spectral luminosity.  Again, for a
\citet{salpeter55} IMF, $\psi$ is a factor of 1.52 larger.  For a
\citet{salpeter55} IMF and $U_\mathrm{SFG} = 1.54 \pm 0.2 \times
10^{19} \mathrm{\,W\,Hz}^{-1}\,\mathrm{Mpc}^{-3}$
(Equation~\ref{eqn:Usfg}), $\psi(t_0) =
0.0128\,M_\odot\,\mathrm{yr}^{-1}\,\mathrm{Mpc}^{-3}$ is the radio
estimate of the SFRD today.

\begin{figure}[!ht]
  \includegraphics[width=0.47\textwidth,trim={4.7cm 9.5cm 6.47cm
      9cm},clip]{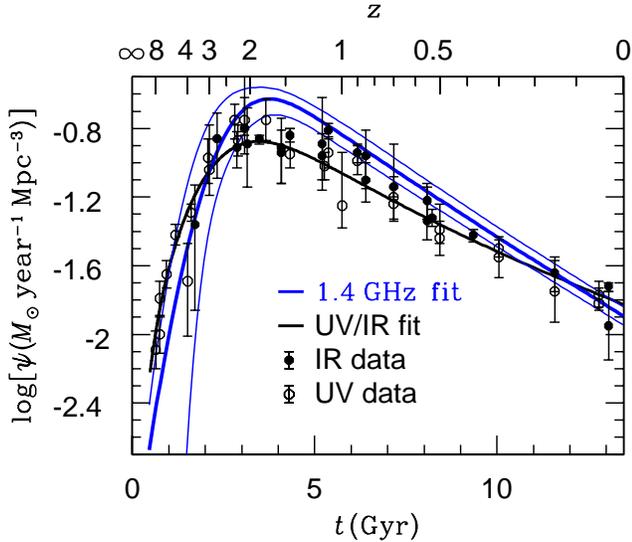}
  \caption{The evolving SFRD $\psi$ for a \citet{salpeter55} IMF is
    shown as a function of time $t$ (Gyr) since the big bang.  The UV
    and IR data points and the black curve fitted to
    Equation~\ref{eqn:madau} with $\psi_0 = 0.015
    \,M_\odot\,\mathrm{yr}^{-1} \,\mathrm{Mpc}^{-3}$ are from the
    \citet{madau14} review.  The heavy blue curve is our best-fit
    1.4\,GHz SFRD estimate, and the light blue curves bound the range
    of acceptable fits to our 1.4\,GHz data.
      \label{fig:psimod}
      }
\end{figure}

Figure~\ref{fig:psimod} compares our 1.4\,GHz estimate of the evolving
SFRD $\psi(t)$ (thick blue curve) with the standard \citet{madau14}
FUV/FIR data points and estimate (black curve), all for a
\citet{salpeter55} IMF. Our best 1.4\,GHz estimate is well
approximated by
%JC I put in the new fit parameters based on your best MCMC fit.
\begin{eqnarray}
  \log \biggl[ \frac {\psi (t)}
    {M_\odot \,\mathrm{yr}^{-1} \, \mathrm{Mpc}^{-3}} \biggr] =
  -3.473 +1.818 \biggl(\frac {t}{\mathrm{Gyr}}\biggr)
  \quad\quad \nonumber \\
  -3.653 \biggl(\frac {t}{\mathrm{Gyr}}\biggr)^2
  +0.02216 \biggl(\frac {t}{\mathrm{Gyr}}\biggr)^3
  \qquad\qquad
\end{eqnarray}
when $0.5 < t\mathrm{(Gyr}) < 5$ and by
\begin{equation}
  \log \biggl[ \frac {\psi (t)}
    {M_\odot \,\mathrm{yr}^{-1} \, \mathrm{Mpc}^{-3}} \biggr] =
          -0.0529  -0.1373   \biggl(\frac {t}{\mathrm{Gyr}}\biggr)
\end{equation}
when $t\mathrm{(Gyr)} > 5$.
The light blue curves indicate the range of SFRDs consistent with the
$13\%$ uncertainty in the SFRD today quadratically added to the SFRD
ranges from our acceptable evolutionary models (Section \ref{sec:SFGevo}).
To convert Figure~\ref{fig:psimod} from a \citet{salpeter55} IMF to a
\citet{kroupa01} IMF, subtract $0.18$ from $\log(\psi)$.  The
sub-linear FIR/radio correlation has increased the 1.4\,GHz late-time
$e$-folding time scale $\tau = 2.9^{+0.07}_{-0.07}\,\mathrm{Gyr}$
(Equation~\ref{eqn:fs}) to $\tau = 3.2^{+0.08}_{-0.08}\,\mathrm{Gyr}$
for the SFRD $\psi$, bringing it closer to but still smaller than the
\citet{madau14} $\tau \approx 4.4\,\mathrm{Gyr}$.
 
\section{Discussion and Conclusions}\label{sec:conclusions}

This paper presents an independent estimate of the cosmic
star-formation history based on radio evolutionary models matching the
1.4\,GHz local luminosity function and counts of sources as faint as
$S = 0.25\,\mu\mathrm{Jy}$ at 1.4\,GHz, the flux density of the Milky
Way at $z = 4$ with $10\times$ luminosity evolution.

\begin{itemize}
  \item Radio source evolution of AGNs and SFGs at 1.4\,GHz was determined by
matching local luminosity functions $\rho_\mathrm{dex}(L_\nu)$ or
local energy-density functions $u_\mathrm{dex}(L_\nu)$ with the
brightness-weighted source counts $S^2 n(S)$.
  \item We made the first measurement of the local \emph{volume-limited}
    FIR/radio correlation and found it to be sub-linear:
    $L_\mathrm{FIR} \propto L_\mathrm{1.4\,GHz}^{0.85}$.
  \item We used our sub-linear FIR/radio correlation to convert radio-source
    evolution to an
evolving SFRD $\psi$ ($M_\odot\,\mathrm{yr}^{-1}\,\mathrm{Mpc}^{-3}$).
This radio estimate reproduces the usual SFRD peak near $z \approx 2$,
but the peak SFRD indicates stronger evolution than the
standard FUV/FIR estimate \citep{madau14}. 
\end{itemize}

\subsection{What are the main
  strengths and weaknesses of this radio SFRD model?}
\begin{itemize}
\item{The 1.4\,GHz emission from a star-forming galaxy is a mixture of
synchrotron radiation from electrons accelerated in core-collapse
supernova remnants of $M > 8 M_\odot$ stars and thermal bremsstrahlung
from H\textsc{ii} regions, making it less sensitive than FIR
luminosity to contamination by older stellar populations. However,
radio emission is more vulnerable to unrecognized AGN contamination,
primarily in galaxies with high SFRs and high radio luminosities.  The
sample of SFGs used to generate the local 1.4\,GHz luminosity function
was carefully vetted \citep{condon19}, and the local radio SFRD
$\psi(t_0) = 0.0128 \,M_\odot\,\mathrm{yr}^{-1} \,\mathrm{Mpc}^{-3}$
is slightly \emph{lower} than the FUV/FIR $\psi(t_0) =
0.015\,M_\odot\,\mathrm{yr}^{-1}\,\mathrm{Mpc}^{-3}$ (both for a
\citet{salpeter55} IMF).  Thus the local 1.4\,GHz sample does not seem
to be badly contaminated.  AGN contamination of SFGs at high redshifts
might cause their source counts and hence radio evolution to be
overestimated, but the excellent agreement of the background
brightnesses $\nu I_\nu$ produced by SFGs at $\nu = 1.4\,$GHz and
$\lambda = 160\,\mu \mathrm{m}$ (Figure~\ref{fig:tbcum2}) is
reassuring.}

\item{AGNs dominate the
source counts above $S \approx 0.4\,\mathrm{mJy}$, but their
contributions to the total counts of significantly fainter sources can
be estimated accurately because they are smooth power laws at the
low-luminosity end of their energy density function. }

  \item{The peak
contribution of SFGs to $S^2 n(S)$ occurs near $S =
10\,\mu\mathrm{Jy}$, and about half the total SFG contribution is from
fainter sources.  The main obstacle to radio measurements of the SFRD
has been measuring accurate source counts down to $S \approx
0.25\,\mu\mathrm{Jy}$.  That is now possible, but only statistically
via the confusion $P(D)$ distribution \citep{matthews21}, so it is not
possible to identify individual $S \approx
0.25\,\mu\mathrm{Jy}$ sources or measure their redshifts.
Instead, the amounts of luminosity evolution $f$ and density evolution
$g$ depend entirely on fitting features in the local energy-density
functions to features in the brightness-weighted source counts.  Only
smoothly varying $f$ and $g$ can be modeled accurately, and rare
populations (e.g., SFGs at very high redshifts) can easily be
overlooked.  The 1.4\,GHz spectra of SFGs are power laws with spectral
indices near $\alpha = -0.7$, so their K-corrections are easy to
calculate but large enough that 1.4\,GHz SFRDs are best determined at
redshifts up to and slightly beyond ``cosmic noon,'' but submm
continuum sources with low or negative K-corrections and submm
spectral lines are better for detecting SFGs at redshifts $z \gtrsim
4$.}

\item{The dominant synchrotron luminosity at 1.4\,GHz is only an
energetically negligible tracer of star formation and is not simply
proportional to the SFR; it depends on unknown or unrelated quantities
such as the interstellar magnetic field strength and inverse-Compton
(IC) scattering off the ambient radiation field produced by starlight
plus the cosmic microwave background (CMB).  Thus the use of 1.4\,GHz
luminosity to measure the SFR is justified primarily by the empirical
FIR/radio correlation.  The locally measured FIR/radio correlation
might fail at high redshifts owing to IC scattering losses off the CMB
$\propto (1 + z)^4$.  This does not seem to be a problem because it
can only lower the radio SFRD estimate, and the radio SFRD estimate is
slightly higher than expected.  The FIR/radio correlation is often
treated as being linear, but we found it to be sub-linear:
$L_\mathrm{FIR} \propto L_\mathrm{1.4\,GHz}^{0.85}$.  Sub-linearity
significantly reduces the discrepancy between the radio and FIR SFRD
models as shown by Figures~\ref{fig:MODS9} and \ref{fig:psimod}, so
the resulting radio SFRD models lie above but just within the error
bars of the FIR data points.}
\end{itemize}
\acknowledgments
We thank the anonymous referee whose insightful and informative
  comments much improved the paper.
The MeerKAT telescope is operated by the South African Radio Astronomy
Observatory, which is a facility of the National Research
Foundation, an agency of the Department of Science and Innovation.
The National Radio Astronomy Observatory is a facility of the National
Science Foundation operated by Associated Universities, Inc.  This
material is based upon work supported by the National Science
Foundation Graduate Research Fellowship under Grant No. DDGE-1315231.
Support for this work was provided by the NSF through the Grote Reber
Fellowship Program administered by Associated Universities,
Inc./National Radio Astronomy Observatory.
This research has made use of the NASA/IPAC Infrared Science Archive,
which is funded by the National Aeronautics and Space Administration and
operated by the California Institute of Technology.\\

\appendix

\section{Gaussian Process Model Fitting}
\label{sec:appendixa}
%JC I reworded this slightly.
Radio source counts and their uncertainties in individual
  flux-density bins are not independent from their neighbors, so
  fitting models to these data by minimizing $\chi^2$ will
  underestimate the model uncertainties and may introduce biases.  We
  use Gaussian processes to allow for possible correlations and derive
  evolutionary models with conservative uncertainties in the
  parameters.

For a complete review of the theory behind (and applications of)
Gaussian processes, we refer the reader to \cite{rasmussen06}.
Briefly, a Gaussian process is a generalization of a Gaussian
probability distribution that takes into account stochastic effects
like correlated noise by modeling both your the function (the physical
model) and also the covariance function.  This ability presents itself
through the generalization of the likelihood function as a matrix
equation
\begin{eqnarray}\label{eq:lnlike}
\log p(\{y_n\}\,|\,\{\bvec{x}_n,\,\sigma_n\},\,\bvec{\theta}) &=&
    -\frac{1}{2}\,{\bvec{r}_\bvec{\theta}}^\T\,K^{-1}\,{\bvec{r}_\bvec{\theta}}
    -\frac{1}{2}\,\log\det K
    -\frac{N}{2}\,\log(2\,\pi)
\end{eqnarray}
where $\bvec{r}_\bvec{\theta}$ is the residual vector
\begin{eqnarray}
{\bvec{r}_\bvec{\theta}}^\T = \left(\begin{array}{ccc}
    y_1 - f(\bvec{x}_1;\,\bvec{\theta}) & \cdots &
    y_N - f(\bvec{x}_N;\,\bvec{\theta})
\end{array}\right)
\end{eqnarray}
and $K$ is the ``covariance matrix.'' When the data points
are independent, the off-diagonal elements of the $N\times N$ matrix $K$
are 0. Covariance between data points $n$ and $m$ are quantified by non-
zero $n,\,m$ off-diagonal elements.
In our case (and in most others) it is difficult or impossible to
estimate the covariances accurately, which makes the ability to fit for
them using Gaussian processes especially helpful. 
 
It would be extremely computationally expensive to add $\sim N^2$ parameters
that
need to be fit. Instead of fitting each $n,\,m$-th element of the matrix
directly, we parameterize it using a functional form
\begin{eqnarray}
K_{n,\,m} &=& {\sigma_n}^2\,\delta_{n,\,m} +
    k(\bvec{x}_n,\,\bvec{x}_m;\,\bvec{\alpha})
\end{eqnarray}
where $\delta_{n,\,m}$ is the Kronecker delta and
$k(\bvec{x}_n,\,\bvec{x}_m;\,\bvec{\alpha})$ is the covariance function
(or kernel) that parameterizes by $\bvec{\alpha})$ the covariance between
by data points using a functional form. It is then up to the user to choose
a covariance function that approximates the (unknown) actual covariance
between data points.

Using the python Gaussian process package \textit{george} \citep{george},
we first maximized the log-likelihood for various covariance functions
to determine which was best suited for our data. We know that the
covariance between data points varies smoothly, and found that the
``squared exponential covariance function'' maximizes the log-likelihood
\begin{eqnarray}
k_\mathrm{SE}(r) = \sigma_f^2\exp\left(-\frac{r^2}{2l^2}\right),
\end{eqnarray}
where $r=|\bvec{x}_n-\bvec{x}_m|$ defines the distance between data points,
$\sigma_r^2$ is a positive constant describing the process variance,
and $l$ defines the characteristic length scale (the reach of influence on
neighboring data points).

\begin{figure}[!htb]
  \centering
  \includegraphics[width=\textwidth]{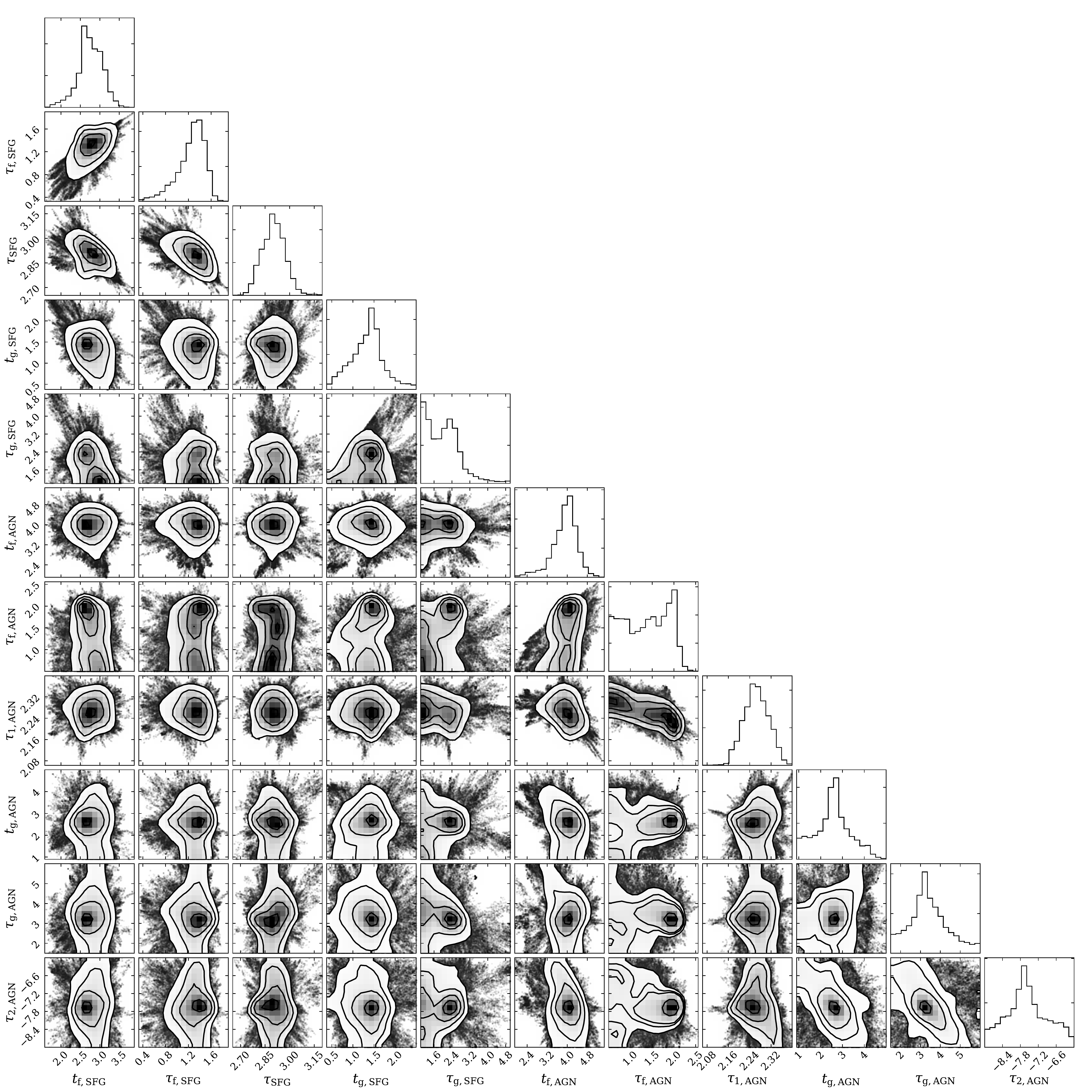}
  \caption{Parameter contours and marginalized posterior distributions
    from the MCMC chains.}
  \label{fig:corner}
\end{figure}

\begin{figure}[!htb]
  \centering
  \includegraphics[trim=2cm 6cm 1cm 11cm, clip, width=0.55\textwidth]{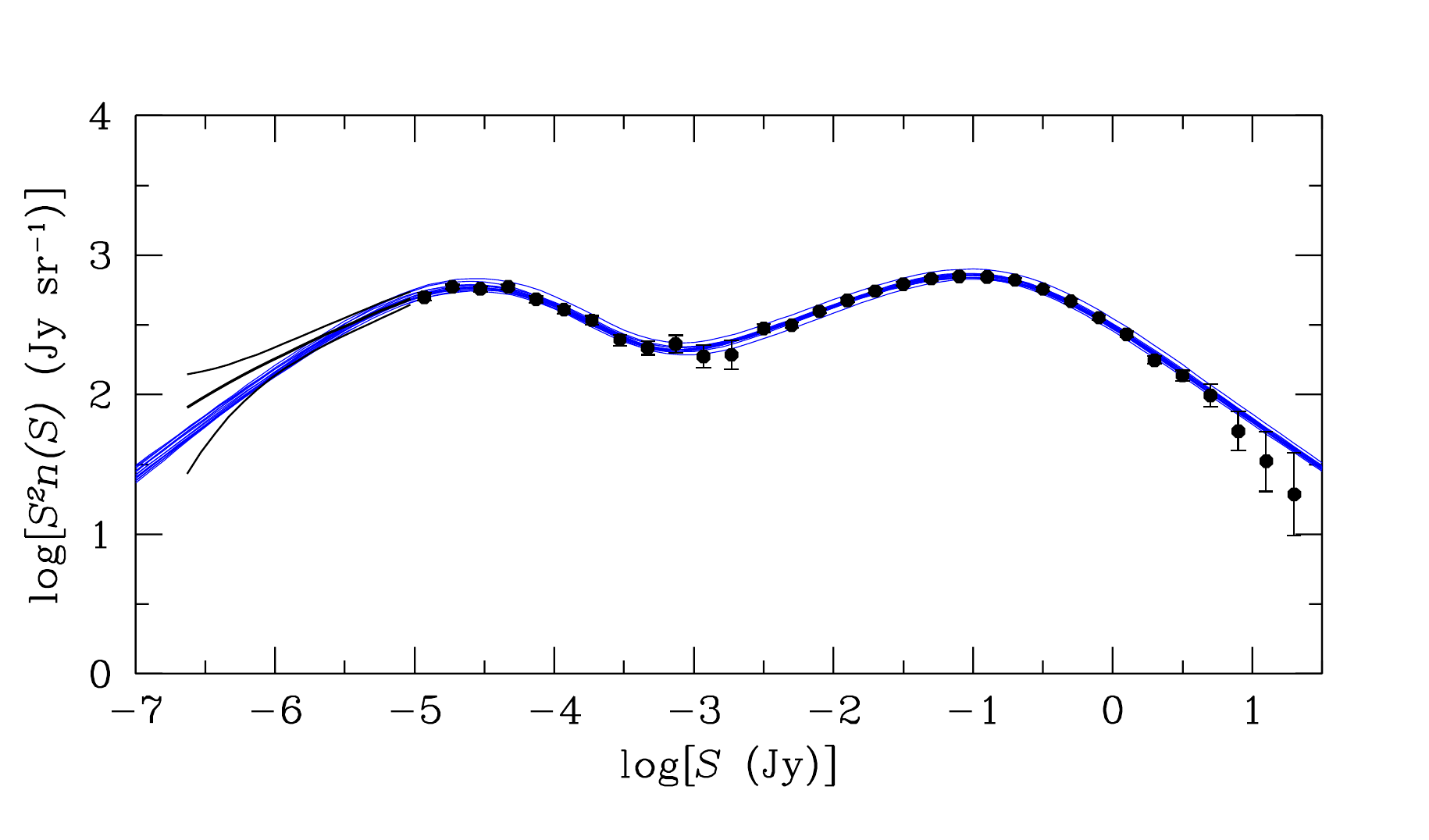}
  \includegraphics[trim=1cm 6.1cm 7cm 11cm, clip, width=0.41\textwidth]{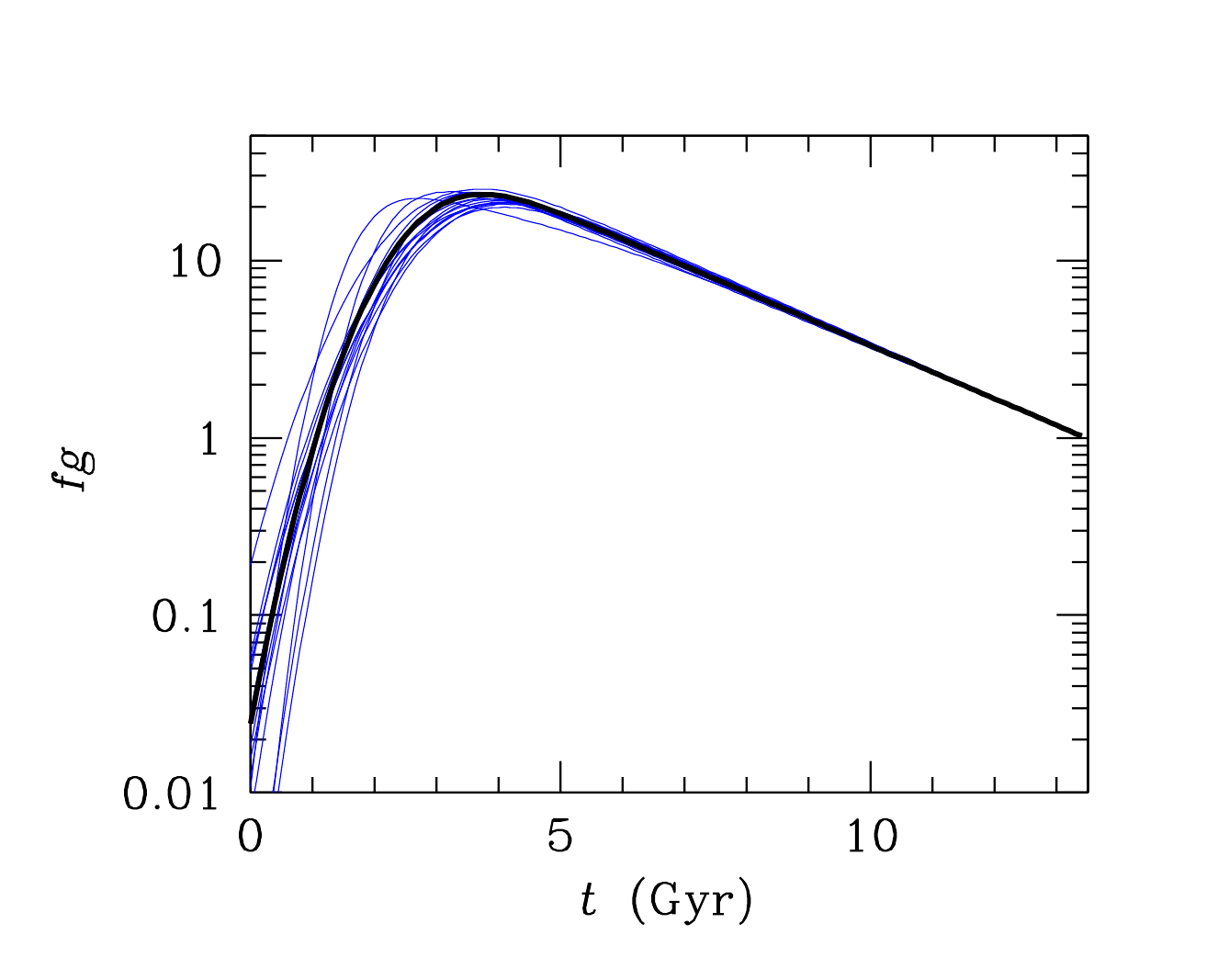}
  \caption{Left: 1.4\,GHz brightness-weighted source count data shown as black points
    (source counts derived via $P(D)$ confusion analysis shown as black curves).
    Fifteen randomly selected parameter vectors from the MCMC fitting routine
    were used to generate predicted source counts (blue curves).
    Right: The corresponding fifteen total evolutionary ($fg$)
    functions for SFGs. The best-fitting total evolutionary
    function is shown as the black solid line.}
  \label{fig:mcmcsamples}
\end{figure}

We used the generalized likelihood function (Equation \ref{eq:lnlike})
with the squared exponential covariance function and the
affine-invariant Markov Chain Monte Carlo code \textit{emcee}
\citep{emcee} to fit for the 13 free parameters: 5 in the equations
governing the evolution of SFGs: $t_\mathrm{f,SFG}
,\,\tau_\mathrm{f,SFG},\,\tau_\mathrm{SFG},\,t_\mathrm{g,SFG}$, and
$\tau_\mathrm{g,SFG}$, 6 in the evolutionary equations for AGNs:
$t_\mathrm{f,AGN}$, $\tau_\mathrm{f,AGN}$, $\tau_\mathrm{1,AGN}$,
$t_\mathrm{g,AGN}$, $\tau_\mathrm{g,AGN}$, and $\tau_\mathrm{2,AGN}$,
plus the two parameters of the covariance function: $\sigma_f^2$
and $l$.  We assumed uniform priors for the input parameters and
enforce the boundary condition $f(0)\cdot g(0) \approx 0$. The
resulting SFG evolutionary parameter contours and marginalized
posterior distributions are shown in Figure \ref{fig:corner}. The
parameter values derived from their marginalized posterior
distributions and their 1$\sigma$ uncertainties are listed in Table
\ref{tab:mcmc}. Source counts resulting from 15 randomly-selected
parameter samples from these posterior distributions and the
corresponding evolutionary functions are shown in Figure
\ref{fig:mcmcsamples}.

\begin{deluxetable}{CcCC}
  \centering
  \tablecaption{MCMC-derived parameter values and uncertainties\label{tab:mcmc}}
  \tablehead{\colhead{Parameter} & \colhead{Best-fit value} & \colhead{$+1\sigma$} & \colhead{$-1\sigma$}}
  \startdata
  t_\mathrm{f,SFG}    &  \phs 2.74  & +0.32 & \phn-0.26\phn \\
  \tau_\mathrm{f,SFG} &  \phs 1.30  & +0.18 & \phn-0.29\phn \\
  \tau_\mathrm{1,SFG} &  \phs 2.90  & +0.07 & \phn-0.07\phn \\
  t_\mathrm{g,SFG}    &  \phs 1.38  & +0.29 & \phn-0.44\phn \\
  \tau_\mathrm{g,SFG} &  \phs 1.99  & +0.67 & \phn-0.76\phn \\
  \hline                                                   
  t_\mathrm{f,AGN}    &  \phs 3.97  & +0.36 & \phn-0.51\phn \\
  \tau_\mathrm{f,AGN} &  \phs 1.41  & +0.54 & \phn-0.65\phn \\
  \tau_\mathrm{1,AGN} &  \phs 2.26  & +0.05 & \phn-0.05\phn \\
  t_\mathrm{g,AGN}    &  \phs 2.59  & +0.75 & \phn-0.89\phn \\
  \tau_\mathrm{g,AGN} &  \phs 3.31  & +1.09 & \phn-0.81\phn \\
  \tau_\mathrm{2,AGN} &  $-$7.62    & +0.84 & \phn-0.67\phn
  \enddata
\end{deluxetable}

\section{The median IR/radio flux-density ratio of faint SFGs}
\label{sec:appendixb}

The median redshift of faint SFGs selected at either $\nu = 1.4$\,GHz
(Section~\ref{sec:tb}) or $\lambda = 160\,\mu$m \citep{berta11} is
$\langle z \rangle \approx 1$, so the observed flux-density ratio
$\langle S_{160\,\mu\mathrm{m}} / S_\mathrm{1.4\,GHz} \rangle$ equals
$\langle S_{80\,\mu\mathrm{m}} / S_\mathrm{2.8\,GHz} \rangle$ in the
source rest frame.  We estimated the latter ratio in terms of the locally
measured quantities FIR (Equation~\ref{eqn:FIRdef}) and $q$
(Equation~\ref{eqn:qdef}).

Nearby SFGs have $\langle q \rangle \approx 2.30$
(Section~\ref{sec:nonlinear}) for flux densities measured at 1.4\,GHz
and SFGs have radio spectral indices $\langle \alpha \rangle \approx
-0.7$ (Section~\ref{sec:basics}), so $ \langle q \rangle \approx 2.51$
for flux densities measured at 2.8\,GHz in the source rest frame.  Local SFGs
typically have FIR flux-density ratios $\langle S_{100\,\mu\mathrm{m}}
/ S_{60\,\mu\mathrm{m}}\rangle \sim 2$ \citep{condon19}, so linear
interpolation in $\log(S), \,\log(\nu)$ between $S_{60\,\mu\mathrm{m}}
\approx 0.68\,\mathrm{Jy}$ and $S_{100\,\mu\mathrm{m}} = 2
S_{60\,\mu\mathrm{m}}$ yields $S_{80\,\mu\mathrm{m}} = 1\,\mathrm{Jy}$
and $\mathrm{FIR} = 3.91 \times 10^{-14} \,\mathrm{W\,m}^{-2}$.  This
result is nearly independent of the ratio $S_{100\,\mu\mathrm{m}} /
S_{60\,\mu\mathrm{m}}$; even for relatively warm SFGs with
$S_{100\,\mu\mathrm{m}} / S_{60\,\mu\mathrm{m}} \sim 1$, the value of
FIR corresponding to $S_\mathrm{80\,\mu\mathrm{m}} = 1\,\mathrm{Jy}$
changes by $<10$\%.  Solving Equation~\ref{eqn:qdef} for
$S_{2.8\,\mathrm{GHz}}$ when $S_{80\,\mu\mathrm{m}} = 1\,\mathrm{Jy}$
gives
\begin{equation}
  \frac {S_{160\,\mu\mathrm{m}}} {S_{1.4\,\mathrm{GHz}}} =
  \frac {S_{80\,\mu\mathrm{m}}} {S_{2.8\,\mathrm{GHz}}} =
  \frac {10^{2.51} \cdot 3.75\times10^{12} \,\mathrm{Hz}} {3.91 \times
     10^{-14}\, \mathrm{W\,m}^{-2} \cdot 10^{26}\,
     \mathrm{Jy\,W}^{-1}\,\mathrm{m}^2 \,\mathrm{Hz}}
        \approx 310\,.
\end{equation}
This flux-density ratio was used to shift the $S_\nu$ and $\nu I_\nu$
axes of Figure~\ref{fig:tbcum2} and demonstrate the excellent
agreement between the observed $\lambda = 160\,\mu\mathrm{m}$ and $\nu =
1.4\,\mathrm{GHz}$ backgrounds produced by SFGs.

A small change in $\langle z \rangle$ has only a small effect
on the calculated ratio $S_{160\,\mu\mathrm{m}} / S_{1.4\,\mathrm{GHz}}$:
\begin{equation}
   \Bigg| d \log \biggl( \frac {S_{160\,\mu\mathrm{m}}} {
S_{1.4\,\mathrm{GHz}}} \biggr) \Bigg|
   <  \big| (\alpha_\mathrm{FIR} - \alpha)  d \log (1+\langle z \rangle)
\big|~.
\end{equation}
For any $1 < S_{100\,\mu\mathrm{m}} / S_{60\,\mu\mathrm{m}} < 2$ and
$0.8 < \langle z \rangle < 1.2$,
  $\log (S_{160\,\mu\mathrm{m}} / S_{1.4\,\mathrm{GHz}})$ varies by less
than $\pm 0.03$.

%\end{document}

%\begin{document}

%\vfil
%\clearpage

\bibliographystyle{aasjournal}
\bibliography{paper2v9}
\vfill\eject
\end{document}